\documentclass[aps,prl,amsmath,amssymb,amsfonts,showpacs,superscriptaddress,floatfix]{revtex4} 
\usepackage{amsmath}
\usepackage{graphicx}
\usepackage{amsthm}

\newtheorem{lemma}{Lemma}

\newcommand{\be}{\begin{equation}}
\newcommand{\ee}{\end{equation}}
\newcommand{\md}{\,{\rm mod}\,}
\begin{document}
\title{Random MERA States and the Tightness of the Brandao-Horodecki Entropy Bound}
\author{M.~B.~Hastings}
\affiliation{Station Q, Microsoft Research, Santa Barbara, CA 93106-6105, USA}
\affiliation{Quantum Architectures and Computation Group, Microsoft Research, Redmond, WA 98052, USA}
\begin{abstract}
We construct a random MERA state with a bond dimension that varies with the level of the MERA.  This causes the state to exhibit a very different
entanglement structure from that usually seen in MERA, with neighboring intervals of length $l$ exhibiting a mutual information proportional to $\epsilon l$ for some constant $\epsilon$, up to a length scale exponentially large in $\epsilon$.  We express the entropy of a random MERA in terms of
sums over cuts through the MERA network, with the entropy in this case controlled by the cut minimizing bond dimensions cut through.
One motivation for this construction is to investigate the tightness of the Brandao-Horodecki\cite{bh} entropy bound relating entanglement
to correlation decay.  Using the random MERA, we show that at least part of the proof is tight: there do exist states with the required property of having linear mutual information between neighboring intervals at all length scales.
We conjecture that this state has exponential correlation decay and that it demonstrates that the Brandao-Horodecki
bound is tight (at least up to constant factors), and we provide some numerical evidence for this as well as a sketch of how a proof of correlation decay might proceed.
\end{abstract}
\maketitle

The amount of entanglement present in a quantum many-body system is closely related to the difficulty of simulating that system and
to the existence of tensor networks to describe that system.  For example, low Renyi entropy in one dimension implies the existence of the
ability to approximate a state by a matrix product state\cite{vc}.

In this regard, an important question has been how much entanglement can be present in a gapped system?  Do such systems obey an area law\cite{eqcp}?  The first general bound showing that a gap implies an area law in one dimension was given in Ref.~\onlinecite{hastings1darea}.  This initial bound gave very poor bounds on the entropy, with the upper bound scaling on the entropy scaling
{\it exponentially} in the local Hilbert space dimension and in the inverse gap.  These results were significantly tightened in Ref.~\onlinecite{tightarea}
 to a scaling that is linear in the inverse gap and polylogarithmic in the local Hilbert space dimension.

Closely connected to this question of entanglement compared to spectral gap is the question of entanglement compared to correlation length.  Indeed, a spectral gap for a local Hamiltonian implies
exponentially decaying correlations\cite{corrdecay} so one might hope to use that correlation decay to prove an entanglement bound.
At a very heuristic level, one might expect that if a system has correlation length $\xi$ and local Hilbert space dimension $D$, then any region $A$ of arbitrary length will only be correlated with degrees of freedom within distance $\xi$ and will be decoupled from the rest of the system.  Let $B$ be the degrees of
freedom within distance $\xi$ of $A$ and let $C$ be the rest of the system.  Then, if $A$ is decoupled from $C$, then $A$ has
entanglement entropy at most roughly $\xi \log(D)$.  This heuristic argument of course has the problem that ``correlations" measure whether there
are operators $O_A,O_C$ supported on $A,C$ such that $\langle O_A O_C \rangle - \langle O_A \rangle \langle O_C \rangle$ is large, while the required decoupling is that $\rho_{AC}$ is close to $\rho_A \otimes \rho_C$ which is a stronger property.

Indeed, early evidence suggested that this heuristic argument was completely incorrect.  Using quantum expanders\cite{qexp1,qexp2}, a family of states were constructed with exponential decay of correlation with uniformly bounded correlation length and fixed local Hilbert space dimension but with arbitrarily high entanglement.  However, more recently in Ref.~\onlinecite{bh}, Brandao and Horodecki showed that exponential correlation length decay {\it did} imply a bound
on entanglement, apparently contradicting the previous result.  The resolution of the apparent contradiction is explained in Ref.~\onlinecite{noteson}.
To explain this resolution, let us first fix some notation.
As in Ref.~\onlinecite{bh}, we define the correlation function between two regions $X,Y$ as
\be
Cor(X:Y)={\rm max}_{\Vert O_X \Vert \leq 1, \Vert O_Y \Vert \leq 1} |{\rm tr}(O_X O_Y \rho)-{\rm tr}(O_X \rho) {\rm tr}(O_Y \rho)|,
\ee
where $O_X,O_Y$ are operators supported on $X,Y$ and $\Vert \ldots \Vert$ denotes the operator norm.  Let us say that a state has $(\xi,l_0,C)$-exponential decay of correlations if for any pair of regions $X,Y$
separated by $l$ sites with $l\geq l_0$ then
\be
Cor(X:Y)\leq C 2^{-l/\xi}.
\ee
In Ref.~\onlinecite{bh}, it is proven that for any connected region $X$, for any pure state on a sufficiently large system which has $(\xi,l_0,1)$-exponential decay of correlations, then
\be
\label{Sbound}
S(\rho_X) \leq c' l_0 \exp(c \log(\xi) \xi),
\ee
for some universal constants $c,c'>0$, where $S(\ldots)$ denotes the von Neumann entropy of a density matrix.  That is, the state obeys an ``area law" as this quantity is independent of the size of $X$; however, it diverges rather rapidly with $\xi$.
Note that this result can also be applied to a system with $(\xi,l_0,C)$-exponential decay of correlations for $C>1$: if a state has
$(\xi,l_0,C)$-exponential decay, then it has $(\xi,\tilde l_0,1)$-exponential decay of correlations, with $\tilde l_0={\rm min}(l_0,\xi \log_2(C)$.

Now we can explain the resolution of the paradox: in Ref.~\onlinecite{bh}, this exponential decay of correlations was assumed to hold for all pairs of
regions $X,Y$.  However, the quantum expander result only shows this exponential decay of correlations for a system on an infinite line, where $X$ represents an interval of sites $[i,j]$ and $Y$ represents another interval of sites $[k,l]$ with $i<j<k<l$.  The quantum expander result would also show
such an exponential decay of a system on a finite line for the same pair of intervals $[i,j]$ and $[k,l]$ if $i$ and $l$ are sufficiently far from the
left and right ends of the line.  
As noted in Ref.~\onlinecite{noteson}, the expander construction does give a $(\xi,l_0,C)$-exponential decay with a $C$ that is uniformly bounded above for such pairs of regions  (although this bound on $C$ was not shown in the original paper applying expanders to
constructing many-body states), so the magnitude of the constant $C$ is not the issue.
However, the quantum expander result does not show correlation decay when $X$ is an interval $[i,j]$ and $Y$ is the union of
a pair of intervals $[k,l]$ and $[m,n]$ with $k<l<i<j<m<n$.  That is, $Y$ is on both the left and right side of $X$, rather than just being to one side.
This difference in the geometry of the regions $X,Y$ considered is the reason for the different result.

So, given the Brandao-Horodecki result, we ask whether this result is tight?  Is
it possible to construct a family of states with $(\xi,l_0,C)$-exponential decay with fixed $C,l_0$ and increasing $\xi$ that have an exponential divergence of
the entanglement with $\xi$?  Rather than building a state using a random expander, we instead turn to a random MERA state\cite{mera}.

While our ultimate goal is to construct a state with large entanglement entropy and small correlations, the Brandao-Horodecki proof provides
some clue as to how to do this.  A key portion of the proof involves considering three regions, called $C,L,R$ (in the proof, they actually write $B_C,B_L,B_R$ to differentiate them from other regions considered, but since we will only consider these three, we just write $C,L,R$) standing for ``center", ``left", ``right".  The center region $C$ is an interval of $2l$ sites, for some $l$.  The left region $L$ consists of the $l$ sites immediately to the left of $C$, while
the right region $R$ consists of the $l$ sites immediately to the right of $C$.
The authors show that if for some choice of regions $C,L,R$ within distance $\exp(1/\xi)$ of $X$ we have $I(C:LR)\leq \epsilon l$ for sufficiently small $\epsilon$, then
the entropy bound (\ref{Sbound}) follows; the authors do this using
the exponential correlation decay to prove the desired area law with the assistance of some results from quantum information theory (the $\epsilon$ needed for this to work depends upon $\xi$, and $\epsilon$ is taken proportional to $1/\xi$). 
Then authors then show that such regions $C,L,R$ do indeed exist: in general for any $\epsilon>0$, then for any site $s$ there are regions $C,L,R$ within $\exp(O(1/\epsilon))$ sites of $s$, with  length $l\leq \exp(O(1/\epsilon))$ such that $I(C:LR)>\epsilon l$.  This is shown using an adaptation of a result in
Ref.~\onlinecite{hastings1darea}.  This is done roughly as follows: suppose that $I(C:LR)>\epsilon l$ for all length scales $l$.
Any interval of a single site has entropy at most $\log(D)$, where $D$ is the Hilbert space dimension on a single site.  So, any interval of length $2$ has entropy at most $2\log(D)$.
Applying the assumption $I(C:LR)$ to $C$ having length $2l=2$, we find that the entropy of an interval of length $4$ is at most $4\log(D)-\epsilon$
and hence any interval of length $8$ has entropy at most $8\log(D)-2\epsilon$.
Then, applying this bound to the case of $C$ having length $2l=8$, we find that the entropy of an interval of length $16$ is at most $16\log(D)-4\epsilon-4\epsilon$.  Iterating this, one eventually finds that the entropy becomes negative at some length scale $\exp(O(1/\epsilon))$ giving a contradiction.

Hence, while we learn from this that we can't achieve $I(C:LR)>\epsilon l$ for all regions, in the state we are constructing we still would like to keep $I(C:LR)$ large (i.e., larger than some constant times $l$) up to some large length scale (i.e., exponentially large in $1/\epsilon$ taking $1/\epsilon$ of order $\xi$) in order to construct a state with
large entanglement entropy and small correlations as if we make $I(C:LR)$ too small, then the area law bound will follow from that step in the Brandao-Horodecki proof.

In fact, it isn't even clear from the proof whether or not it is possible to construct even a state such that $I(C:LR)\geq\epsilon l$ for all regions $C,L,R$ with $l$ sufficiently small compared to $\exp({\rm const.}/\epsilon)$ for some constant which may depend upon the Hilbert space dimension on each site.  If this were not possible (for example, if one could only have $I(C:LR)\geq \epsilon l$ for $l$ small compared to $1/\epsilon$) this would immediately tighten the Brandao-Horodecki result.  So, part of our construction will be showing that this is possible.
We will in fact show a mutual information lower bound that implies this one: we will construct a state such that for every pair of neighboring intervals of
length $l\leq \exp(1/\epsilon)$, the mutual information is lower bounded by $\epsilon l$ (in fact, since we have a random state, we will show this result in expectation; see the Discussion).  This will have the side effect of producing mutual information between $C$ and $LR$ in the choice of intervals above: the left half of $C$ will have mutual information with $L$ and the right half of $C$ will have mutual information with $R$.  

Note it is easy to construct a state such that this bound holds for very particular choices of $C,L,R$.  That is, we can ensure that the bound holds for at least one choice of $C,L,R$ at each length scale up to $\exp({\rm const.}/\epsilon)$ as follows.  We now sketch this construction.  The tools we develop later to analyze the MERA construction can be used to analyze this case and verify the claims in this paragraph.
Construct a quantum circuit with the form of a binary tree.  The input to the quantum circuit is a state in a $1$-dimensional Hilbert space.  The node at the top of the tree
represents an isometry that maps this state to a system of two sites, each with some given dimension $D_1$ for some $D_1$; i.e., this is an isometry from a $1$-dimensional Hilbert space
to a $D_1 \times D_1$-dimensional Hilbert space.  The two nodes at the next level of the tree represent further isometries, mapping each $D_1$-dimensional Hilbert space to a pair of
$D_2$-dimensional Hilbert spaces, and so on.  We choose these isometries at random, and we choose the dimensions $D_k$ for $k>1$ so that $D_k$ is slightly larger than $\sqrt{D_{k-1}}$;
more accurately, take $\log(D_{k})\approx (1/2)\log(D_{k-1})+\epsilon 2^{L-k}$, where $L$ is the total number of levels of the tree and $\epsilon$ is some small number.
The leaves of the tree represent the final states of the system.  Call two nodes ``siblings" if they are both at the same level of the tree and have the same parents.
One may guess then from the increase in dimension of the Hilbert space that there will be some mutual information between the leaves which are descendants of any given node and those which are descendants of the sibling of that node.  The entropy of the descendants of the first node will be roughly $\log(D_k)$ if the node is at level $k$, as will be the entropy of the descendants of the other node, while the entropy of the combination will be only $\log(D_{k-1}) <2\log(D_k)$.  However, for other choices of intervals of leaves of the tree, there will be almost no mutual
information.

However, this construction only gives the mutual information for certain choices of $C,L,R$.  We would like to have it holds for all choices.  So, we instead use a MERA state.
This is similar to the tree state, except with additional ``disentanglers".

Our construction of a random MERA state has some properties that may have a holographic interpretation.  See the Discussion.

An important question is whether the state that we construct indeed has exponential decay of correlations.  We conjecture that it does and we sketch
how a proof of such conjecture might proceed and we provide some numerical evidence.  However, we leave a proof of this statement for the
future.

\section{MERA state}
\begin{figure}
\includegraphics[width=8cm]{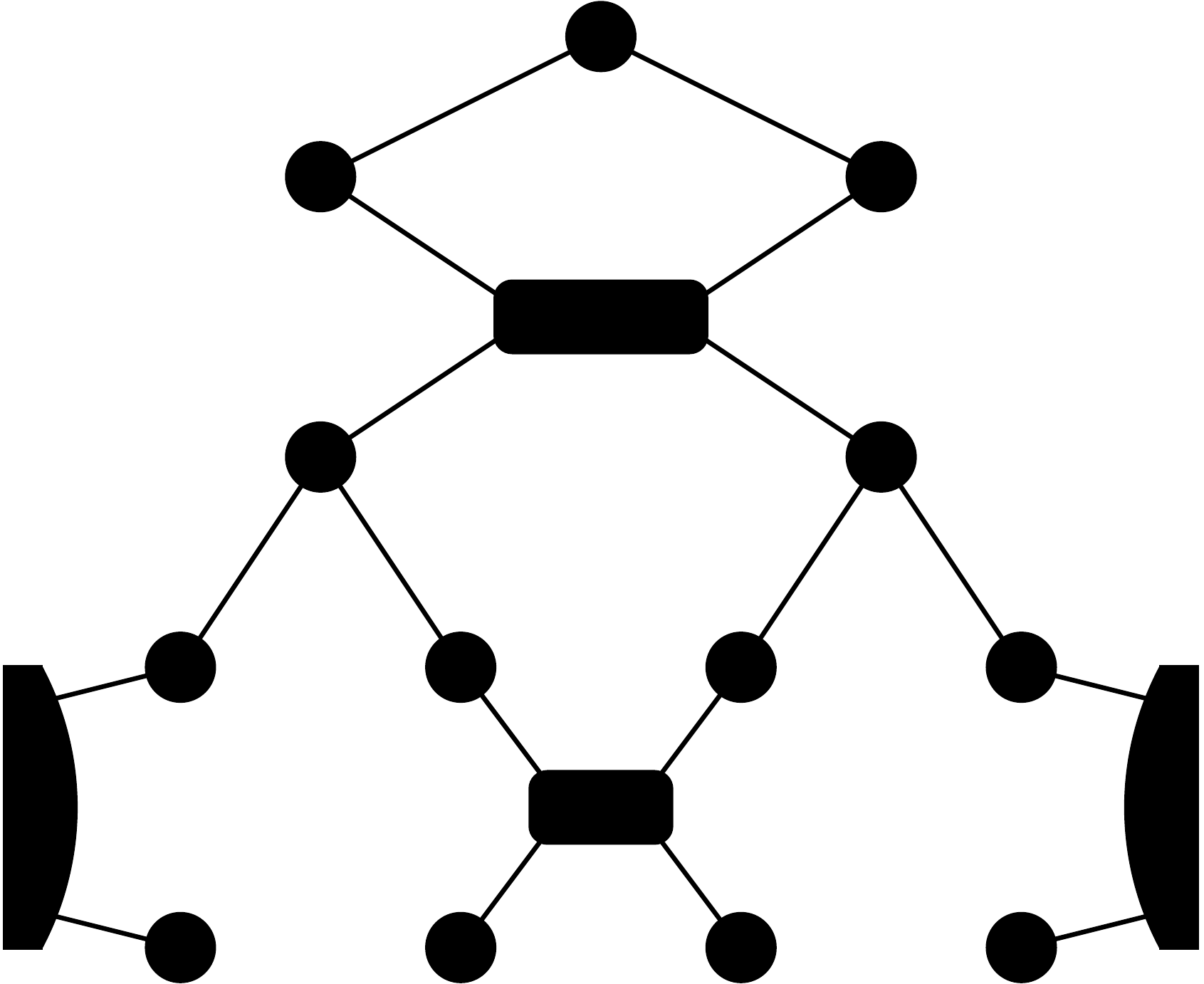}
\caption{Illustration of MERA network.  Circle at top represents state $\psi_0$.  Isometry $V_1$ is represented by the lines leading to a pair of circles below it.  Isometry $W_1$ is represented by the filled rectangle mapping that pair of circles to another pair of circles (note that in this case, $W_1$ could be absorbed into a redefinition of $V_1$, while $W_i$ for $i>1$ cannot be absorbed into $V_i$).  Isometry $V_2$ maps each circle in the pair to another pair of circles.  Isometry $W_2$ maps the four sites to another four sites.  The isometry on sites $1,2$ is represented by the filled rectangle in the middle, while the isometry on sites $0,3$ is represented by the lines leading to half a filled rectangle on left and right sides of the figure.}
\label{MERAFig}
\end{figure}

We define the MERA network as follows.  See Fig.~\ref{MERAFig}.
We start with a single site with a $1$-dimensional Hilbert space (thus, up to an irrelevant choice of phase, the state of the system on this site is fixed; call this initial state $\psi_0$).
We then apply a series of isometries to this state, giving a new state
\be
\label{psidef}
\psi=W_L V_L \ldots W_3 V_3 W_2 V_2 W_1 V_1  \psi_0,
\ee
for some $L$, where $L$ is the  number of ``levels".  The final state $\psi$ is a state on $N=2^{L}$ sites.
Each $V_k$ is an isometry that maps a system on $2^{k-1}$ sites with some Hilbert space dimension $D_{k-1}$ on each site to a system on $2^{k}$ sites with some Hilbert space dimension $D'_k$ on each site.
We number the sites before applying $V_k$ by numbers $0,1,2,...,2^{k-1}$ and after applying $V_k$ by numbers $0,1,2,...,2^{k}-1$.  
Each $V_k$ is a product of isometries on each of the $2^{k-1}$ sites, mapping each site to a pair of sites; the $j$-th site
is mapped to a pair of sites $2j,2j+1$.
Each $W_k$ is another isometry.  The isometry $W_k$ preserves the number of sites, mapping a system of $2^{k}$ sites with dimension $D'_k$ on each site to a system of $2^{k}$ sites with dimension $D_{k}$ on each site.
Each $W_k$ is also a product of isometries, but in this case it is a product of isometries on pairs of sites; it maps each pair $2j+1,2j+2 \md 2^{k}$ to the same pair.

We will say that isometries $W_i$ with smaller $i$ are at {\it higher} levels of the MERA while those with larger $i$ are at {\it lower} levels of the MERA.  That is, the {\it height} of a level
will increase as we move upwards in the figure.  Each ``level" of the MERA will include two rows of the figure, one with the isometry $W$ and one with the isometry $V$.

Note that pairs of sites are defined modulo $2^{k}$ in the definition of $W_k$.  If the sites are written on a line in order $0,...,2^{k}-1$, then $W_k$ will entangle the rightmost and leftmost sites.  The introduction of $W_1$ in the definition of $\Psi$ above is slightly redundant, since $V_1$ already produces entanglement between sites $0,1$; however, we leave $W_1$ in to keep the definition of the MERA consistent from level to level.

We will explain the choice of dimensions $D_k,D'_k$ later.  In a difference from traditional MERA states, the dimensions $D_k,D'_k$ will be chosen differently at each level.  Further, the dimension $D_{k}$ will be larger than $D'_k$.  That is, the $W_k$ (sometimes called ``disentanglers") will have the effect of increasing the Hilbert space dimension of each site, and hence of the system as a whole.

The isometries $W_k,V_k$ will be chosen randomly.  More precisely, each $V_k$ is product of isometries on each of the $2^k$ sites, mapping each site to
a pair of sites. Each of the isometries in this product will be chosen at random from the Haar uniform distribution, independently of all other isometries.  Similarly, each $W_k$ is also a product of isometries, each of which will again be chosen at random from the Haar uniform distribution, independently of all other isometries.

\section{Entanglement Entropy of Interval}
We now estimate the entanglement entropy of an interval of sites.  We start with some notation.  We write $[i,j]$ to denote the interval of sites $i,i+1,...,j-1,l$.
We define
$\psi(k)=W_{k} V_{k} ... W_1 V_1 \psi_0$, so that $\psi=\psi(L)$ and we define $\sigma(k)=|\psi(k)\rangle\langle \psi(k)|$.
We define $\phi(k)=V_{k} W_{k-1} V_{k-1} ... W_1 V_1 \psi_0$ and we define $\tau(k)=|\phi(k)\rangle\langle\phi(k)|$.
We begin with an upper bound to the von Neumann entropy using a recurrence relation.
We then derive a similar recurrence relation for the expectation value of the second Renyi entropy and use that to lower bound the expected von Neumann entropy.
We then combine these bounds to get an estimate on the expected entropy of an interval.
These general bounds will hold for any sufficiently large choice of $D_k,D'_k$; we then specialize to a particular choice to obtain the desired state with large entanglement.

\subsection{Upper Bound to von Neumann Entropy By Recurrence Relation}
We begin with a trivial upper bound for $S(\sigma(k)_{[i,j]})$, which denotes the von Neumann entropy of the reduced density matrix of $\sigma(k)$ on the interval
$[i,j]$.
Since the $W_k$ are isometries, we have
\be
\label{isomresultW}
i \md 2=1,j \md 2=0 \; \rightarrow \; S(\sigma(k)_{[i,j]})=S(\tau(k)_{[i,j]}).
\ee
That is, in the case that $i$ is odd and $j$ is even, the interval $[i,j]$ in state $\sigma(k)$ is obtained by an isometry acting on that interval in the state $\tau(k)$.
If $i=j$, we have the bound
\be
\label{ieqjbound}
i = j \; \rightarrow \; S(\sigma(k)_{[i,j]}) \leq D_k.
\ee
In all other cases (if, for example $i$ is even or $j$ is odd or both), the entropy can be bounded above using subadditivity:
\be
\label{subaddbound}
S(\sigma(k)_{[i,j]})\leq S(\sigma(k)_{[m,n]})+\log(D_k) (|m-i| + |n-j|),
\ee
for any choice $m,n$.
Combining Eqs.~(\ref{isomresultW},\ref{subaddbound}) gives us the bound
\be
\label{svnsigmabound}
S(\sigma(k)_{[i,j]}) \leq {\rm min}_{m,n \, s.t. \, |m-i| \leq 1, |n-j| \leq 1}^{m \md 2=1,n\md 2=0} \Bigl( S(\tau(k)_{[m,n]})+\log(D_k) (|m-i| + |n-j|) \Bigr).
\ee
Although in fact this equation holds for any choices of $m,n$, for all applications we will restrict to $m,n$ such that  $|m-i|\leq 1,|n-j| \leq 1$.

We emphasize that in the above equation, and from now on, all differences, such as $m-i$, are taken modulo the number of sites at the given level of the MERA.  When we compute a difference such as $m-i$, by $|m-i|$ we mean
the integer $k$ with minimum $|k|$, such that $m-i = k$ modulo the number of sites.  Similarly, if we write, for two sites, $i,j$ that $i=j+1$, we again mean modulo the number of sites at the given level.

Of course, if we have the empty interval, which we write as $[i,j]$ for $i=j+1 \md 2^k$, then the entropy is equal to $0$.

Similarly we have
\be
\label{svntaubound}
S(\tau(k)_{[i,j]}) \leq 
{\rm min}_{m,n \, s.t. \, |m-i| \leq 1, |n-j| \leq 1}^{m \md 2 =0, n \md 2=1} 
\Bigl( S(\sigma(k-1)_{[m/2,(n-1)/2]})+\log(D'_{k}) (|m-i| + |n-j|) \Bigr).
\ee
We will only use Eq.~(\ref{svntaubound}) with $|m-i| \leq 1, |n-j| \leq 1$.

\subsection{Expectation Value of Renyi Entropy}
We now obtain a recurrence relation for the expectation value of the Renyi entropy $S_2$, defined by $S_2(\rho)=-\log({\rm tr}(\rho^2))$.
The analogue of Eq.~(\ref{isomresultW}) still holds for $S_2$:
\be
\label{S2isomresultW}
i \md 2=1,j \md 2=0 \; \rightarrow \; S_2(\sigma(k)_{[i,j]})=S_2(\tau(k)_{[i,j]}),
\ee
as does
\be
\label{reduce}
S_2(\sigma(k)_{[i,j]}) \leq {\rm min}_{m,n \, s.t. \, |m-i| \leq 1, |n-j| \leq 1}^{m \md 2=1,n\md 2=0} \Bigl( S_2(\tau(k)_{[m,n]})+\log(D_k) (|m-i| + |n-j|) \Bigr)
\ee
and
\be
\label{reduce2}
S_2(\tau(k)_{[i,j]}) \leq 
{\rm min}_{m,n \, s.t. \, |m-i| \leq 1, |n-j| \leq 1}^{m \md 2 =0, n \md 2=1} 
\Bigl( S_2(\sigma(k-1)_{[m/2,(n-1)/2]})+\log(D'_{k}) (|m-i| + |n-j|) \Bigr).
\ee
We refer to Eqs.~(\ref{reduce},\ref{reduce2}) as the {\it reduction equations}.
These equations make sense also in the case that we have $m>n$.  This can occur, for example, if $j=i$ or $j=i+1$ in which case the equations
allow us to bound $S_2(\sigma(k)_{[i,j]}) \leq \log(D_k) |j-i+1|$ and $S_2(\tau(k)_{[i,j]}) \leq \log(D'_k) |j-i+1|$

We now show that the upper bound given by repeatedly applying these equations to obtain the optimum result (i.e., the result which minimizes the $S_2$)
is tight for the expectation value of $S_2$, up to some corrections proportional to a certain level in the MERA.  
That is, we give a lower bound on the expectation value of $S_2$.
Consider first $S_2(\sigma(k)_{[i,j]})$.  Assume first that $i \neq j$ and
$i\md 2=0,j \md 2=0$ (we discuss the other cases later; they will be very analogous to this case).
We write the Hilbert space of the system of sites $0,...,2^{k}-1$ as a tensor product of four Hilbert spaces.  These will be labelled $A_1,A_2,B,R$ where $A_1$ is the Hilbert space on site $i-1$, $A_2$ is the Hilbert space site $i$,
$B$ is the Hilbert space on sites $i+1,...,j$, and $R$ is the Hilbert space on all other sites.  In this case, the Hilbert space on a set of sites
refers to the case in which there is a $D_k$-dimensional Hilbert space on each site.
In this notation, $S_2(\sigma(k)_{[i,j]})=S_2(\sigma(k)_{A_2 B})$.
The isometry $W_{k}$ is a product of isometries on pairs of sites.  Write $W_{k}=W X$ where $W$ is the isometry acting on the pair of sites $i-1,i$ and $X$ is the product of all other isometries.
The isometry $W_{k}$ maps from a system of $2^k$ sites to a system of $2^k$ sites, but it changes the Hilbert space dimension from $D'_{k}$ to $D_k$.
We introduce different notation to write the Hilbert space of the system with a $D'_{k}$-dimensional space on each site.  We write it is a product of spaces $a,b,r$, where $a$ is the Hilbert space for sites $i-1,i$,
$b$ is the Hilbert space on sites $i+1,...,j$, and $r$ is the Hilbert space on all other sites.
Then,
\be
S_2(\sigma(k))_{A_2 B})=S_2\Bigl({\rm tr}_{A_1 r}(W \tau(k) W^\dagger)\Bigr).
\ee
That is, we wish to compute the entanglement entropy of $W\phi(k)$ on $A_2 b$.  The isometry $W$ is from $a$ to $A_1 \otimes A_2$.

\begin{figure}
\includegraphics[width=10cm]{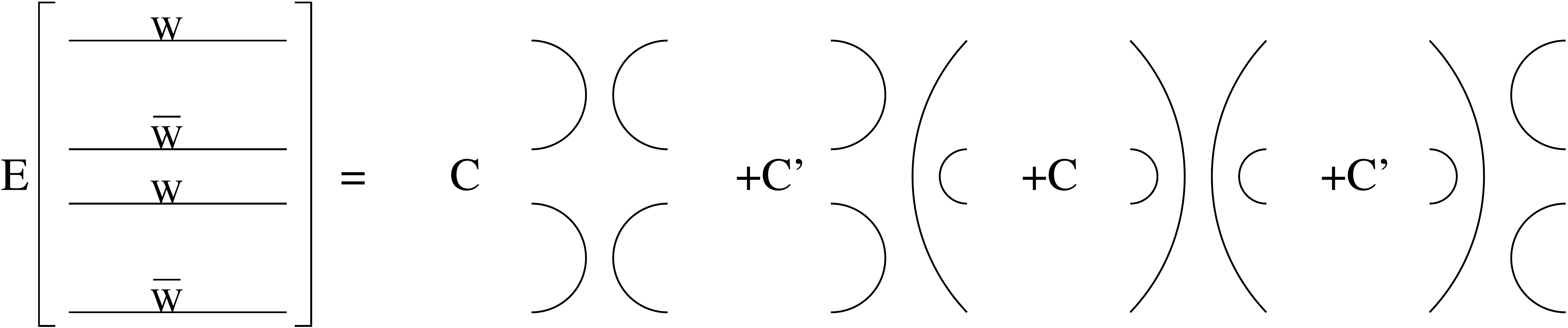}
\caption{Identity for average of $W$.  Left-hand side represents expectation value of a product of two powers of $W$ and two powers of $\overline W$.  Right-hand side pictorially shows the result of Eq.~(\ref{identity}), where the arcs represents Kronecker $\delta$.}
\label{WidFig}
\end{figure}

Note that since the logarithm is a concave function and so the negative of the logarithm is a convex function, we have
\begin{eqnarray}
\label{convexity}
E\Bigl[S_2\Bigl({\rm tr}_{A_1 r}(|W \tau(k) W^\dagger)\Bigr)\Bigr]_W &=& -E\Bigl[\log {\rm tr}\Bigl([{\rm tr}_{A_1 r}(W \tau(k) W^\dagger)]^2\Bigl)\Bigr]_W \\ \nonumber
&\geq &  -\log E\Bigl[{\rm tr}\Bigl([{\rm tr}_{A_1 r}(W \tau(k) W^\dagger)]^2\Bigl)\Bigr]_W,
\end{eqnarray}
where $E[\ldots]_W$ denotes the average over $W$.
The trace ${\rm tr}\Bigl([{\rm tr}_{A_1 r}(W \tau(k) W^\dagger)]^2\Bigl)$ is a second-order polynomial in $W$ and second-order polynomial in the complex conjugate of $W$.  For an arbitrary isometry $W$ from a Hilbert space of dimension $d_1$ to a Hilbert space of dimension $d_2$ (in this case, $d_1=(D'_{k})^2$ and $d_2=(D_k)^2$ since $W$ is an isometry from pairs of sites to pairs of sites; note also that $d_2 \geq d_1$ since this is an isometry),
we can average this trace over choices of $W$ using the identity for the matrix elements of $W$ and $\overline W$:
\begin{eqnarray}
\label{identity}
&&E[W_{ij} \overline W_{kl} W_{ab} \overline W_{cd}]_W \\ \nonumber
&=& c\Bigl( \delta_{ik} \delta_{jl} \delta_{ac} \delta_{bd}+\delta_{ic} \delta_{jd} \delta_{ka}\delta_{lb} \Bigr) \\ \nonumber
&&+c'\Bigl( \delta_{ik} \delta_{jd} \delta_{ac} \delta_{lb} + \delta_{ic} \delta_{jl} \delta_{ka} \delta_{bd} \Bigr),
\end{eqnarray}
where
\be
\label{cis}
c=\frac{d_1^2 \cdot(d_1^2d_2^2+d_1d_2)-d_1\cdot(d_1^2d_2+d_1d_2^2)}{(d_1^2d_2^2+d_1d_2)^2-(d_1^2d_2+d_1d_2^2)^2},
\ee
and
\be
\label{cprimeis}
c'= \frac{d_1\cdot(d_1^2d_2^2+d_1d_2)-d_1^2\cdot(d_1^2d_2+d_1d_2^2)}{(d_1^2d_2^2+d_1d_2)^2-(d_1^2d_2+d_1d_2^2)^2}
\ee
Eq.~(\ref{identity}) is illustrated in Fig.~\ref{WidFig}.
Some of these averages are similar to calculations in Ref.~\onlinecite{hw}.

Note that the right-hand side of Eq.~(\ref{identity}) is the most general function that is invariant under unitary rotations $W \rightarrow U W U'$ for arbitrary unitaries $U,U'$ and invariant under interchange $i,j \leftrightarrow a,b$ or $k,l \leftrightarrow c,d$.  The constants $c,c'$ can be fixed by taking traces with $\delta_{ik} \delta_{jl} \delta_{ac} \delta_{bd}$ and 
 $\delta_{ik} \delta_{jd} \delta_{ac} \delta_{lb}$ and computing the expectation value.  The trace of the right-hand side with $\delta_{ik} \delta_{jl} \delta_{ac} \delta_{bd}$
is equal to $c(d_1^2d_2^2+d_1d_2)+c'(d_1^2d_2+d_1d_2^2)$.  One can readily show that the trace with the left-hand side is equal to
$d_1^2$, as the trace with $\delta_{ik} \delta_{jl} \delta_{ac} \delta_{bd}$ is independent of the choice of $W$.
The trace of the right-hand side with  $\delta_{ik} \delta_{jd} \delta_{ac} \delta_{lb}$ is equal to $c'(d_1^2d_2^2+d_1d_2)+c(d_1^2d_2+d_1d_2^2)$, while the trace with the left hand side is equal to $d_1^2$.
So, this gives
\be
d_1^2=c(d_1^2d_2^2+d_1d_2)+c'(d_1^2d_2+d_1d_2^2),
\ee
\be
d_1=c'(d_1^2d_2^2+d_1d_2)+c(d_1^2d_2+d_1d_2^2).
\ee
Solving these gives Eqs.~(\ref{cis},\ref{cprimeis}).

If we use Eq.~(\ref{identity}) to compute $E\Bigl[{\rm tr}\Bigl([{\rm tr}_{A_1 r}(W \tau(k) W^\dagger)]^2\Bigl)\Bigr]_W$, we find a sum of four terms, one for each of
the terms on the right-hand side of Eq.~(\ref{identity}).  The result is
\begin{eqnarray}
\label{resultS2av}
&&(c+c')D_k^3\Bigl({\rm tr}\Bigl([{\rm tr}_{a r}(\tau(k))]^2\Bigr)+{\rm tr}\Bigl([{\rm tr}_{r}(\tau(k))]^2\Bigl)\Bigr)
\\ \nonumber
&=&
\frac{D_k^3}{D_k^4+D_k^2}\Bigl({\rm tr}\Bigl([{\rm tr}_{a r}(\tau(k))]^2\Bigr)+{\rm tr}\Bigl([{\rm tr}_{r}(\tau(k))]^2\Bigl)\Bigr)
\\ \nonumber
&=& \frac{1}{D_k}\cdot(1-O(1/D_k))\cdot \Bigl({\rm tr}\Bigl([{\rm tr}_{a r}(\tau(k))]^2\Bigr)+{\rm tr}\Bigl([{\rm tr}_{r}(\tau(k))]^2\Bigl)\Bigr),
\end{eqnarray}
where the asymptotic $O(...)$ notation refers to scaling in $D_k$ in this equation.

Note that since both terms on the right-hand side of last line of Eq.~(\ref{resultS2av}) are positive, the last line is at most equal to twice the maximum term.

Note that the terms on the right-hand side are related to Renyi entropy; for example, minus the logarithm of 
${\rm tr}\Bigl([{\rm tr}_{a r}(\tau(k))]^2\Bigl)$ is equal to $S_2(\tau_{ar}(k))$, i.e. the $S_2$ Renyi entropy of $\tau(k)$ on $ar$, and similarly for $r$.
So, we get:
\be
\label{logsumexp}
E[S_2(\sigma(k)_{[i,j]})]_W \geq -\log\Bigl\{\sum_{m,n \, s.t. \, |m-i| \leq 1, |n-j| \leq 1}^{m \md 2=1,n\md 2=0} \exp\Bigl(-S_2(\tau(k)_{[m,n]})-\log(D_k) (|m-i| + |n-j|) \Bigr)\Bigr\}.
\ee
The reader might note that we have in fact only derived Eq.~(\ref{subaddboundavg}) in the case that $j \md 2=0$.  However, by averaging over isometries at both left and right ends of the interval, one can handle the case that $j \md 2=1$ identically to above.

Note that also that since all terms in the sum of Eq.~(\ref{logsumexp}) are positive, the sum is bounded by a constant times the maximum.  So,
minus the logarithm of the right-hand side is equal to
$$
\log(D_k)+{\rm max}\Bigl(S_2(\tau_{ar}(k)))+S_2(\tau_{r})\Bigr)-O(1).$$
So, using Eq.~(\ref{convexity}), we have that
\be
\label{subaddboundavg}
E[S_2(\sigma(k)_{[i,j]})]_W \geq {\rm min}_{m,n \, s.t. \, |m-i| \leq 1, |n-j| \leq 1}^{m \md 2=1,n\md 2=0} \Bigl( S_2(\tau(k)_{[m,n]})+\log(D_k) (|m-i| + |n-j|) \Bigr)-O(1).
\ee
Here the $O(1)$ notation refers to a term bounded by a constant, independent of all dimension $D_k,D'_k$.

We will also want the result that
\be
\label{exprecursigma}
E[\exp\Bigl(-S_2(\sigma(k)_{[i,j]})\Bigr)]_W \leq \sum_{m,n \, s.t. \, |m-i| \leq 1, |n-j| \leq 1}^{m \md 2=1,n\md 2=0} \exp\Bigl(-S_2(\tau(k)_{[m,n]})-\log(D_k) (|m-i| + |n-j|) \Bigr).
\ee
Note that the left-hand side of the above equation is the quantity
$E\Bigl[{\rm tr}\Bigl([{\rm tr}_{A_1 r}(W \tau(k) W^\dagger)]^2\Bigl)\Bigr]_W$
that we have been considering and Eq.~(\ref{logsumexp}) follows from this by convexity.

We also give the analogs of Eq.~(\ref{logsumexp},\ref{subaddboundavg},\ref{exprecursigma}) for the entropy of $\tau(k)$:
\be
\label{logsumexptau}
E[S_2(\tau(k)_{[i,j]})]_W \geq 
-\log\Bigl\{{\rm sum}_{m,n \, s.t. \, |m-i| \leq 1, |n-j| \leq 1}^{m \md 2 =0, n \md 2=1} 
\exp\Bigl(-S_2(\sigma(k-1)_{[m/2,(n-1)/2]})-\log(D'_{k}) (|m-i| + |n-j|) \Bigr)\Bigr\},
\ee
\be
\label{subaddboundavgtau}
E[S_2(\tau(k)_{[i,j]})]_W \geq 
{\rm min}_{m,n \, s.t. \, |m-i| \leq 1, |n-j| \leq 1}^{m \md 2 =0, n \md 2=1} 
\Bigl( S_2(\sigma(k-1)_{[m/2,(n-1)/2]})+\log(D'_{k-1}) (|m-i| + |n-j|) \Bigr)-O(1),
\ee
\be
\label{exprecurtau}
E[\exp\Bigl(-S_2(\tau(k)_{[i,j]})\Bigr)]_W \leq 
{\rm sum}_{m,n \, s.t. \, |m-i| \leq 1, |n-j| \leq 1}^{m \md 2 =0, n \md 2=1} 
\exp\Bigl(-S_2(\sigma(k-1)_{[m/2,(n-1)/2]})-\log(D'_{k-1}) (|m-i| + |n-j|) \Bigr).
\ee

This now allow us to upper and lower bound the expectation value of $S_2$ for any interval $[i,j]$.  Given an interval $[i,j]$, let a {\it reduction sequence} denote a sequence of choices at each level to reduce $[i,j]$ to the empty interval so that at each step we apply either Eq.~(\ref{reduce}) or Eq.~(\ref{reduce2}) until we are left with $i>j$ at which point we are left with the empty interval which has entropy $0$.  That is, such a sequence consists first of a choice $m,n$
with $m \md 2=1,n\md 2=0$, followed by a choice $m,n$ with $m \md 2 =0, n \md 2=1$, and so on, with $i,j$ at each step being determined by the $m,n$ at the previous step.
For such a sequence $Q$, let $S(Q)$ denote the upper bound to the entropy obtained from the reduction equations; in this reduction, once we obtain the empty interval, we use the fact that that has entropy $0$.
Let $h(Q)$ denote the {\it height} of a given reduction sequence, namely the number of times we apply the reduction equations until we arrive at the empty interval.  Note that the height increases by $2$ every time we change the level by $1$ since we apply both equations.

Then, we have the result that
\be
S_2(\tau_{[i,j]})\leq {\rm min}_Q S(Q),
\ee
and
\be
\label{seqsum}
E[\exp(-S_2(\tau_{[i,j]}))] \leq \sum_Q \exp(-S(Q)),
\ee
as follows by using Eqs.~(\ref{exprecursigma},\ref{exprecurtau}) to sum $\exp(-S_2(\ldots))$ over reduction sequences.
As a point of notation, from now on we use $E[...]$ to denote the average over all $W,V$ in the MERA.

Finally, we have
\begin{lemma}
\label{lbndlemma}
\be
\label{lbnd}
E[S_2(\tau_{[i,j]})] \geq {\rm min}_Q S(Q)-O(1) h(Q).
\ee
\begin{proof}
From Eq.~(\ref{seqsum}) and convexity of $-\log(\ldots)$,
\be
E[S_2(\tau_{[i,j]})] \geq \log\{\sum_Q \exp(-S(Q))\}.
\ee
Write the sum over $Q$ inside the logarithm as a sum over levels,
\be
\sum_Q \exp(-S(Q))=\sum_h \sum_{Q, h(Q)=h}  \exp(-S(Q)).
\ee
Since there are at most $4^Q$ sequences of height $h$ (we have at most two choices at each side of the sequence,
\be
\sum_{Q, h(Q)=h}  \exp(-S(Q)) \leq {\rm max}_{Q,h(Q)=h} \exp(-S(Q)+\ln(4) h(Q)).
\ee
Now, we use a general identity.  Let $g(x)$ be any positive function such that $\sum_{x=1,2,\ldots} g(x)^{-1}$ converges to some constant $c$.
Then, for any function $f(x)$, we have that $\sum_{x=1,2,\ldots} f(x) \leq c \cdot {\rm max}_{x=1,2,\ldots} f(x) g(x)$.
To verify this identity, minimize $c \cdot {\rm max}_{x=1,2,\ldots} f(x) g(x)$ over positive functions $f$ subject to a constraint on $\sum_{x=1,2,\ldots} f(x)$; the minimum will be attained for $f(x)$ proportional to $1/g(x)$ and plugging in this choice of $f(x)$ gives the identity.
So, picking $g(x)=(1/2)^x$, we get that
\begin{eqnarray}
&& \sum_h {\rm max}_{Q,h(Q)=h} \exp(-S(Q)+\ln(4) h(Q))  \\ \nonumber
& \leq & {\rm max}_h {\rm max}_{Q,h(q)=h} \exp(-S(Q)+\ln(8) h(Q)) \\ \nonumber
&=& {\rm max}_{Q}  \exp(-S(Q)+\ln(8) h(Q)).
\end{eqnarray}
Then, Eq.~(\ref{lbnd}) follows, choosing the $O(1)$ constant to be $\log(8)$.
\end{proof}
\end{lemma}

We remark (we will not need this for this paper) that for some choices of $D_k,D'_k$, the sum in Eq.~(\ref{seqsum}) will be
dominated by a single reduction sequence.  In that event, it will be possible to tighten Eq.~(\ref{lbnd}) by improving on the term $-O(1)h(Q)$ on the
right-hand side.

Further, we also have
\begin{lemma}
\label{vnineq}
The following inequalities for the von Neumann entropy hold:
\be
\label{l1}
S(\tau_{[i,j]})\leq {\rm min}_Q S(Q),
\ee
\be
\label{l2}
E[S(\tau_{[i,j]})] \geq {\rm min}_Q S(Q)-O(1) h(Q).
\ee
\begin{proof}
Eq.~(\ref{l1}) holds 
by the reduction equations (\ref{svnsigmabound},\ref{svntaubound}) for $S$, 
and Eq.~(\ref{l2}) follows by lemma \ref{lbndlemma} since $S$ is greater than $S_2$.
\end{proof}
\end{lemma}

\section{Choice of $D_k,D'_k$}
We now give the choice of $D_k,D'_k$.  At the bottom of the MERA state, the leaves have dimension $D_L$ chosen to be any fixed value greater than $1$.  For example, we may take $D_L=2$.  Then, we would follow the recursion relations:
\be
\label{apprecur1}
\log(D'_k) \approx \log(D_{k})-\epsilon 2^{L-k},
\ee
for all $k$
and for $k<L$
\be
\label{apprecur2}
\log(D_{k})\approx 2\log(D'_{k+1})-\epsilon 2^{L-k}.
\ee
The value $\epsilon$ here will be related to the $\epsilon$ in the Brandao-Horodecki paper and to the mutual information that we find between intervals.
The factor of $2^{L-k}$ represents the length scale associated to a given level in the MERA state: there are roughly $2^{L-k}$ leaves of the MERA in the future light-cone of a given
node at level $k$.  Here, the ``future light-cone" refers to the leaves such that there is a path in the MERA starting at the given node at level $k$ and moving downward, ending at the given leaf.  The usual terminology in MERA instead refers to a {\it causal cone} of operators being mapped upwards to higher levels of the MERA; we discuss this later.

We write the approximation symbol $\approx$ rather than the equals symbol $=$ because the dimensions $D_k,D'_k$ should be integers.
So, the recursion relations we use to obtain integer dimensions are
\be
D'_k=\lceil \exp\{      \log(D_{k})-\epsilon 2^{L-k}    \} \rceil,
\ee
\be
D_{k} =\lceil \exp\{  2\log(D'_{k+1})-\epsilon 2^{L-k}     \} \rceil.
\ee
We choose $\epsilon,L$ so that $D_0=1$.
This can be done taking $L \sim 1/\epsilon$ so that the total number of sites in the system is equal to $\exp(\Theta(1/\epsilon))$.
The calculation is essentially that in Ref.~\onlinecite{hastings1darea} and Ref.~\onlinecite{bh}, where both papers used a recursion relation
for the entropy.  Let us study the recursion relations ignoring the complications of the ceiling; that is, we treat Eqs.~(\ref{apprecur1},\ref{apprecur2}) as if
they were exact.  The ceiling in the correct recursion relations has negligible effect on the scaling behavior.
We have $D_L$ given.  Then, $\log(D_{L-1})=2\log(D_L)-3\epsilon$.  Then, $\log(D_{L-2})=4\log(D_L)-12\epsilon$ and $\log(D_{L-3})=8\log(D_L)-36\epsilon$.  In general,
\be
\label{Dkest}
\log(D_{L-m}) \approx 2^m\log(D_L)-3 m\epsilon 2^{(m-1)}.
\ee
This remains positive until $m \sim 1/\epsilon$; so, as claimed, we can take $L \sim 1/\epsilon$.
Note also that for all $m<L-1$,
\be
\log(D_{L-m}) \gtrsim 2^m \epsilon.
\ee
We say that for all $m<L-1$ because $\log(D_1)$ must be positive, so $\log(D_2)$ must be at least $\epsilon 2^{L-2}$; for many choices of $\epsilon,L$,
as similar inequality will hold even for $m=L-1$ and we will always choose $\epsilon,L$ such that this holds.

\subsection{Entanglement Entropy For This Choice}
We now estimate the entanglement entropy for this choice of $D_k,D'_k$ for an interval $[i,j]$.
We make a remark on the Big-O notation that we use.  When we say in lemma \ref{intentlemma} and lemma \ref{milemma} that a quantity
is $\Omega(x)$, we mean that it is lower bounded by $c_1x-c_2\log(l)-c_3$ for some positive constants $c_1,c_2,c_3$ which do not depend on $D_L,\epsilon$.  We emphasize this because otherwise one might worry about subleading terms hidden in the Big-O notation: since the leading term often involves a factor of $\epsilon$ (at least in lemma \ref{milemma}), a quantity such as $\epsilon l$ becomes large only once $l$ becomes large enough and so
one might worry about the simultaneously limits of large $l$ and small $\epsilon$.  The notation $O(1)$ continues to refer to a quantity bounded
by a constant, independent of $\epsilon,D_L$.

\begin{lemma}
\label{intentlemma}
The expected entanglement entropy of an interval $[i,j]$ with length $l=j-i+1$ with $l \neq N/2$ is lower bounded by
\be
\label{intentlower}
E[S(\tau_{[i,j]})] \geq \Omega(\log(D_{L-\log_2(l)})).
\ee
\begin{proof}
We estimate ${\rm min}_Q S(Q)-O(1) h(Q)$ and apply lemma \ref{vnineq}.  For any choice of $[i,j]$, for any sequence $Q$, each time we apply
Eq.~(\ref{reduce}) or Eq.~(\ref{reduce2}), it is possible that we produce a positive  term, $\log(D_k) (|m-i| + |n-j|)$ or $\log(D'_{k-1}) (|m-i| + |n-j|)$, respectively.  Let us say that if $|m-i|=1$ then the term is applied at the ``left end" of the interval, while if $|n-j|=1$, then the term is applied at the right end of the interval (as the interval changes as we change level in the MERA by applying Eqs.~(\ref{reduce},\ref{reduce2}), we continue to define the left end and right end in the natural way).

One may verify that at least every other time we apply the equations, we must produce a positive term at the left end and 
 at least every other time we apply the equations, we must produce a positive term at the right end.  That is, if Eq.~(\ref{reduce}) does not produce a positive term at the left (or right) end, then Eq.~(\ref{reduce2}) must produce a positive term at the left (or right, respectively) end.
The only exception to this is if the interval becomes sufficiently long that it includes all sites at the given level of the MERA; this does not happen for the intervals considered here.
So,
\be
\label{SQlower}
S(Q)\geq 2(D_L+D_{L-1}+D_{L-2}+\ldots+D_{L-\lfloor h(Q)/2 \rfloor}).
\ee

We now estimate the minimum $h(Q)$.  Every time we apply Eq.~(\ref{reduce}) and then Eq.~(\ref{reduce2}), an interval of length $l$ turns into an interval of length at least $l/2-2$.  The factor of $-2$ occurs because Eq.~(\ref{reduce}) can reduce the length by at most $2$; then Eq.~(\ref{reduce2}) can further reduce the length by at most $2$ more, and then divide the length by $2$.
Thus, $2$ applications of this pair of equations can map an interval of length $l$ to one of length $(l/2-2)/2-2=l/2-3$, and $k$ applications can map an interval of length $l$ to one of length $l/2^k-4$.  Once the length becomes four or smaller, than the length can be mapped to zero by a pair of applications.  Thus, $h(Q)$ is greater than or equal to $2k$ with $l/2^{k}-4 \leq 4$, so $l/2^{k}\leq 8$ (in fact this estimate is not quite tight, as if the length of the interval is nonzero after $k$ applications, then $h(Q)>2k$).  So,
\be
\label{hQlower}
h(Q) \geq 2 \log_2(l)-O(1).
\ee

Combining Eqs.~(\ref{SQlower},\ref{hQlower}) gives Eq.~(\ref{intentlower}).
Here we use Eq.~(\ref{Dkest}) to estimate $D_{L-h(Q)}$ and note that $\log(D_{L-\log_2(l)+O(1)}) \geq \Omega(\log(D_{L-\log_2(l)}))$.
Further, we use the fact that the term $-O(1) h(Q)$ in $S(Q) - O(1) h(Q)$ is asymptotically negligible compared to $S(Q)$.
\end{proof}
\end{lemma}

\subsection{Mutual Information}
We now estimate the mutual information between a pair of neighboring intervals, each of length $l$.  We lower bound this by
$\epsilon l$.  This implies a similar lower bound on the mutual information between a single interval $[i,j]$ of length $2l$ and its two neighboring intervals of length $l$.
\begin{lemma}
\label{milemma}
The expected mutual information between two neighboring intervals $[i,j]$ and $[j',k]$, with $j'=j+1$ and $l=j-i+1=k-j$ is lower bounded by
\be
\label{milower}
E[I([i,j];[j',k])] \geq \Omega(\epsilon l).
\ee
\begin{proof}
Call $[i,j]$ the ``left interval" and call $[j',k]$ the ``right interval".
Let $Q_L, Q_R$ be reduction sequences for $[i,j]$  and $[j',k]$, respectively, which minimize $S(Q)-O(1)h(Q)$.
So, $E[S(\tau_{[i,j]})+S(\tau_{[j',k]})] \geq S(Q_L)+S(Q_R)-O(1) h(Q_L)-O(1)h(Q_R)$.

We now show that
$S(\tau_{[i,k]}) \leq S(Q_L)+S(Q_R)-\Omega(\epsilon) l$.  
Note that always the optimum reduction sequences have $h(Q_L), h(Q_R) \leq {\rm const}. \times \log(l))$.  So, this upper bound on $S(Q)$
will imply Eq.~(\ref{milower}).  This bound will be based on constructing a reduction sequence for $[i,l]$; however, we will in one case need to
also use subadditivity and then construct further reduction sequences.  That is, it will not simply be a matter of applying Eqs.~(\ref{reduce},\ref{reduce2}) with a given sequence but a more general reduction procedure will be needed.

Let $i$ be the left end of interval $[i,j]$ and $j$ be the right end.
Refer to Fig.~\ref{redcurveFig}.
The reduction sequence $Q_L$ describes how both the left and right end of the left interval move as we change levels in the MERA.  Let $i_0,i_1,i_2,...,i_{h(Q_L)}$ be the sequence describing where
the left end is after each application and let $j_0,...,j_{h(Q_L)}$ describe where the right end is.  That is,
after $k$ applications of Eqs.~(\ref{reduce},\ref{reduce2}), we have a new interval $[i_k,j_k]$.  Eventually, after $h(Q_L)$ applications of the reduction
equation,the interval has length zero so that $i_{h(Q_L)}=j_{h(Q_L)}+1$.
Similarly, let $j'_0,...,j'_{h(Q_R)}$ and $k_0,...,k_{h(Q_R)}$ be the left and right ends of the right interval.

Let $S_L(Q_L)$ denote the sum of the quantities
 $\log(D_k)|m-i|$ or $\log(D'_{k-1}) |m-i|$ obtained using  Eqs.~(\ref{reduce},\ref{reduce2}) for reduction sequence $Q_L$ 
, while let $S_R(Q_L)$ denote the sum of 
 $\log(D_k)|n-j|$ or $\log(D'_{k-1}) |n-j|$.  That is, these are the sum of the terms at the left or right ends of the interval, so that
$S(Q_L)=S_L(Q_L)+S_R(Q_L)$.  Define $S_L(Q_R)$ and $S_R(Q_R)$ similarly so $S(Q_R)=S_L(Q_R)+S_R(Q_R)$.

Suppose first that $i_a=k_a$ for some given $a$, i.e., the left end of $Q_L$ meets the right end of $Q_R$.  Then,
define a reduction sequence $Q$ by taking the sequence $i_0,...,i_a$ for the left end of $Q$ and $k_0,...,k_a$ for the right end of $Q$.
Then, $S(\tau_{[i,l]}) \leq S(Q)\leq S_L(Q_L)+S_R(Q_R)=S(Q_L)+S(Q_R)-S_R(Q_L)-S_L(Q_R)$.  However, referring to the calculation in lemma \ref{intentlemma},
$S_R(Q_L)\geq \Omega(\epsilon l)$ as is $S_L(Q_R)$, which gives the desired result.

So, let us assume that $i_a \neq k_a$ for all $a$.  Suppose, without loss of generality that $h(Q_L) \geq h(Q_R)$.  In fact, it may not be possible that $h(Q_L)$ will ever differ from $h(Q_R)$ for the optimal
sequences $Q_L,Q_R$ for the given pair of intervals and for the given choice of dimensions in the network so it might suffice to always assume that
$h(Q_L)=h(Q_R)$, but we are able to lower bound the mutual information even in this
possibly hypothetical case (it is possible for $h(Q_L)$ to differ from $h(Q_R)$ if the intervals have different length).
 
To simplify notation, let $h=h(Q_R)$.
Define $B_L(Q_L)$ to be the sum over the first $h$ applications 
of Eqs.~(\ref{reduce},\ref{reduce2}) in reduction sequence $Q_L$ of
 $\log(D_k)|m-i|$ or $\log(D'_{k-1}) |m-i|$, while let $B_R(Q_L)$ denote the sum over the first $h$ applications of
 $\log(D_k)|n-j|$ or $\log(D'_{k-1}) |n-j|$.  The notation $B_L$ or $B_R$ is intended to indicate that these are the contributions to $S_L$ or $S_R$ arising from the first $h$ applications, i.e., at the ``bottom" of the MERA.

Consider applying Eqs.~(\ref{reduce},\ref{reduce2}) a total of $Q_R$ times, using the sequence $i_0,...,i_{h}$ for the left
end and $k_0,...,k_{h}$ for the right end.  Note that this sequence of reductions may not end at the empty interval; rather, it leaves the interval
$[i_{h},k_{h}]$.
This gives
\be
\label{firstbound}
S(\tau_{[i,k]}) \leq B_L(Q_L)+S_R(Q_R)+\Delta,
\ee
where $\Delta$ is either 
\be
\Delta=S(\tau(L-h/2)_{[i_{h},k_{h}]})
\ee
if $h$ is even or
\be
\Delta=S(\sigma(L-(h-1)/2)_{[i_{h},k_{h}]})
\ee
if $h$ is odd.
That is, $\Delta$ is the entropy of the interval that remains after applying the reduction sequence.

\begin{figure}
\includegraphics[width=8cm]{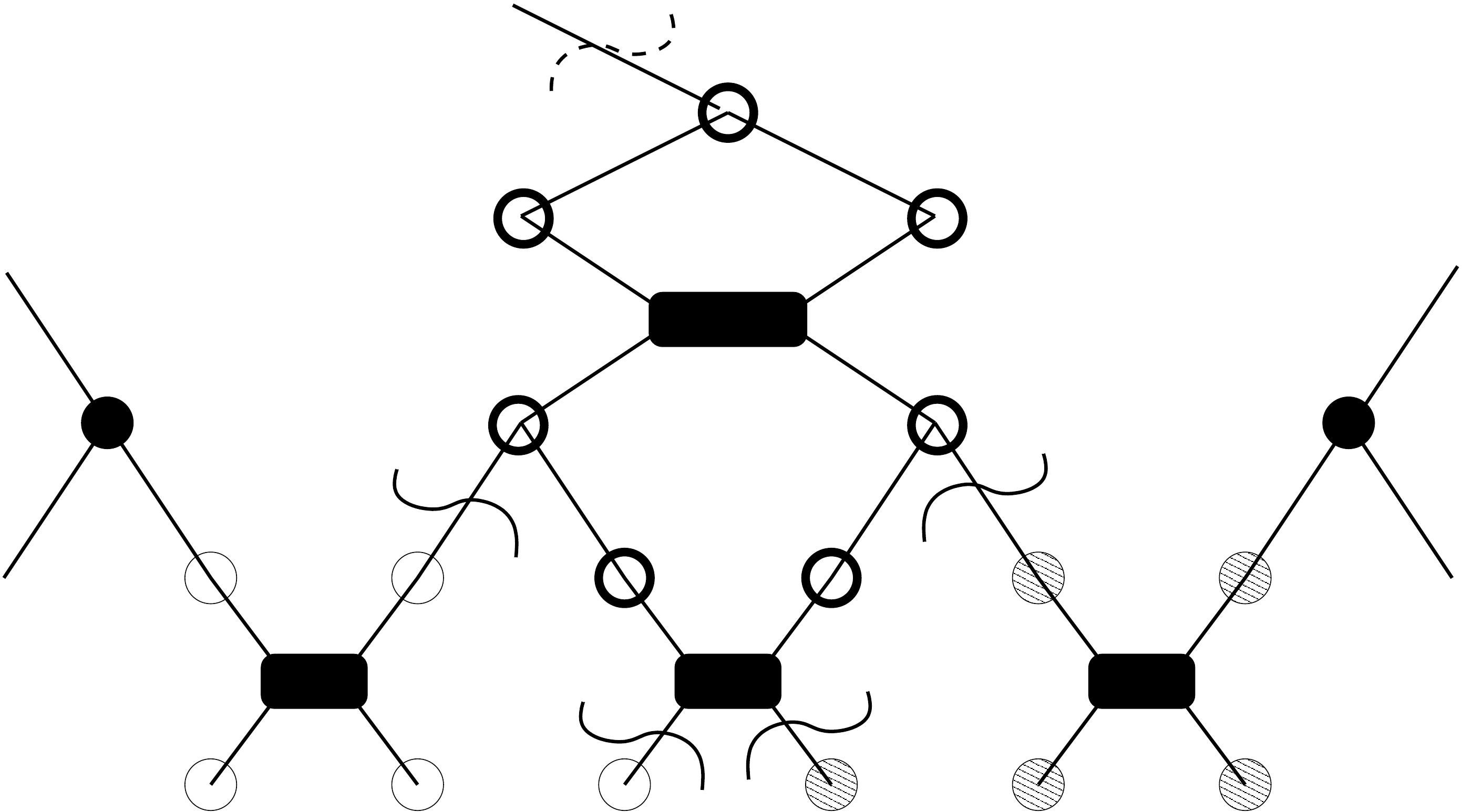}
\caption{Illustration of part of a MERA network.  Only a fragment of the network is shown, so that the three lines leaving upwards connect to other parts of the network, as do the two lines leaving downwards. We illustrate computing mutual information between two intervals, each of three sites.  The left interval is represented by the unfilled circles on the leaves of the
MERA, while the right interval is represented by the circles with diagonal lines.  Unfilled circles with thin outer lines and circles with diagonal lines at higher levels of the MERA represent the intervals that
result from applying Eqs.~(\ref{reduce},\ref{reduce2}) for the optimum sequences $Q_L,Q_R$, respectively.  Both sequences have $h=2$.  When computing the optimum reduction of
the $6$-site interval containing both of these $3$-site intervals, the resulting intervals contain the unfilled circles with thin outer lines, and the circles with diagonal lines and also the
unfilled circles with thicker outer lines.  Filled circles indicate sites not in any of these reduction sequences.  The squiggly lines crossing the lines of the MERA network represent contributions to $S_R(Q_L)$ and $S_L(Q_R)$, while the dashed
squiggly line crossing the line at the top represents an extra term in the entropy to reduce the $6$-site interval.  The difference between these is equal to the expectation value
of the mutual information, up to subleading terms.}
\label{redcurveFig}
\end{figure}

Now use subadditivity.  To simplify notation, let us suppose that $h$ is even (we simply do this so that we can write $\tau(...)$ everywhere, rather
than having to specify either $\tau(...)$ or $\sigma(...)$ in each case).
Then,
\be
\label{split}
\Delta \leq S(\tau(L-h/2)_{[i_h,j_h]}) + S(\tau(L-h/2)_{[j_h+1,k_h]}).
\ee
Note that $k_h=j'_{h}-1$ since the right interval vanishes after $h$ applications of the reduction equations; this makes the interval $[j_h+1,k_h]$ look more symmetric in left and right.  Note also that if $h(Q_L)=h(Q_R)$, then $j_h+1=i_h$.

We can then upper bound $S(\tau(L-h/2)_{[i_h,j_h]})$ using a reduction sequence with left end $i_h,...,i_{h(Q_L)}$ and right end $j_h,...,j_{h(Q_L)}$, giving
\be
S(\tau(L-h/2)_{[i_h,j_h]}) \leq S_L(Q_L)-B_L(Q_L) + S_R(Q_L)-B_R(Q_L).
\ee
So, from Eqs.~(\ref{firstbound},\ref{split}),
\be
S(\tau_{[i,k]}) \leq S_L(Q_L)+\Bigl( S_R(Q_L)-B_R(Q_L) \Bigr) + S_R(Q_R)+S(\tau(L-h/2)_{[j_h+1,k_h]}).
\ee

However,
\be
S(\tau(L-h/2)_{[j_h+1,k_h]}) \leq B_R(Q_L)+S_L(Q_R)-\Omega(\epsilon l),
\ee
which gives the desired bound on the mutual information.
To see this, estimate 
$S(\tau(L-h/2)_{[j_h+1,k_h]})$ using another reduction sequence.  
In Fig.~\ref{redcurveFig}, the interval $[j_h+1,k_h]$ consists of the two sites with open circles on the row two rows above the bottom (i.e., the bottom row of the level one level
above the bottom).
Then entropy 
$S(\tau(L-h/2)_{[j_h+1,k_h]})$ is less than or equal to $(k_h-j_h)*\log(D_{L-h/2})$.  However, $B_R(Q_L)+S_L(Q_R)$
is greater than or equal to $(k_h-j_h)*\log(D_{L-h/2})$ as can be seen in the figure \ref{redcurveFig}; that is, the entropy of the two sites $[j_h+1,k_h]$ is greater than or equal to
the sum of logarithms of dimensions of bonds cut by squiggly lines.
If $k_h-j_h$ is sufficiently large, then in fact $S(\tau(L-h/2)_{[j_h+1,k_h]}) \leq (k_h-j_h)*\log(D_{L-h/2}) - \Omega(\epsilon l)$; this simply requires that $k_h-j_h$ be large enough that
at least one pair of sites in the interval $[j_h+1,k_h]$ emerge from same isometry as occurs in the figure.
Alternately, if $k_h=j_h+1$, then $S(\tau(L-h/2)_{[j_h+1,k_h]})\leq \log(D_{L-h})$, while $B_R(Q_L)+S_L(Q_R)\geq 2\log(D_{L-h})$.
If $k_h \leq j_h$, then 
$S(\tau(L-h/2)_{[j_h+1,k_h]})=0$.
The remaining case is the $k_h=j_h+2$ but that the two sites do not emerge from the same isometry.  However, in this case $S(\tau(L-h/2)_{[j_h+1,k_h]})\leq 2 \log(D_{L-h/2})$.  
However, by the same calculation in lemma \ref{intentlemma} that gave Eq.~(\ref{SQlower}), we find that
$B_R(Q_L)\geq D_L+D_{L-1}+D_{L-2}+\ldots+D_{L-h/2}$ and
$S_L(Q_R)\geq D_L+D_{L-1}+D_{L-2}+\ldots+D_{L-h/2}$, so 
$B_R(Q_L)+S_L(Q_R) -S(\tau(L-h/2)_{[j_h+1,k_h]}) \geq 2D_{L-h/2+1}+\ldots$ which is $\Omega(\epsilon l)$.
\end{proof}
\end{lemma}

\section{Correlation Decay}
We now discuss decay of correlations in this state.  We do not prove correlation decay.  However, we conjecture
that
for the MERA state above,
for any two regions $X,Y$ separated by distance $l$, we have
$Cor(X:Y)\leq C 2^{-l/\xi}$ for some $C=O(1)$ and some $\xi$ bounded by $O(1)/\epsilon$ with probability that tends to $1$ as $D_{L-\log_2(l)}$ tends
to infinity.
In the discussion, we briefly discuss controlling rare events.

The simplest version of this correlation decay to consider is when $X$ consists of a single site and $Y$ is separated from $X$ by at least $1$ site.
Thus, the site in $X$ consists of one of the two sites which is in the output of some given isometry $W$, and $Y$ does not contain the other site
which is in the output of that isometry.
Let us divide the system into three subsystems.  Let $B=X$.  Let $E$ be the other site which is in the output of the same isometry as $X$, and
let $A$ consist of the rest of the system.  (We rename $X$ as $B$ to make the notation more suggestive of a quantum channel from Alice to Bob,
as we will use ideas from quantum channels.)
Since $Y\subset A$, it suffices to consider correlation functions $\langle \psi | O_A O_B |\psi \rangle$ for $O_A,O_B$ supported on $A,B$ respectively.

Consider the two subsystems $A$ and $BE$, and
make a singular value decomposition of the wavefunction $\psi$, so that we write
\be
\psi=\sum_{\alpha} A(\alpha) |\alpha\rangle_A \otimes |\alpha\rangle_{BE},
\ee
where $|\alpha\rangle_A$ and $|\alpha\rangle_{BE}$ are complete bases of states on $A$ and $BE$, respectively, and $A(\alpha)$
are complex scalars with $\sum_{\alpha} |A(\alpha)|^2=1$.
Let $O_A$ have matrix elements$(O_A)_{\beta,\alpha}$ in this basis.
Then,
\begin{eqnarray}
&& \langle \psi i | O_A O_B |\psi \rangle \\ \nonumber
&=&\sum_{\alpha,\beta} \overline{A(\beta)} A(\alpha) \Bigl( {}_A\langle \beta | O_A | \alpha \rangle_A \Bigr)
\Bigl( {}_{BE} \langle \beta | O_B | \alpha \rangle_{BE} \Bigr) \\ \nonumber
&=& {\rm tr}(\tilde O_A O_B),
\end{eqnarray}
where $\tilde O_A$ is defined by its matrix elements
\be
(\tilde O_A)_{\beta\alpha}=(O_A)_{\alpha\beta}\overline{A(\beta)} A(\alpha).
\ee

To estimate the correlation decay, we must maximize the correlation function over $O_A,O_C$ with $\Vert O_A \Vert,\Vert O_C \Vert\leq 1$.
Since the maximization over operators with bounded infinity norm may not be easy, we instead derive a bound
in terms of a maximization over operators with bounded $\ell_2$ norm (which we write $|\ldots|_2$) for which the maximization reduces to a problem in linear algebra.
We have
\begin{eqnarray}
\label{weakbound}
|\tilde O_A|_2 &\equiv &\sqrt{{\rm tr}(\tilde O_A^2)} \\ \nonumber
&=&\sqrt{\sum_{\beta\alpha} |A(\beta)|^2 |A(\alpha)|^2|(O_A)_{\beta\alpha}|^2} \\ \nonumber
& \leq & \Bigl( {\rm max}_{\alpha} |A(\alpha)|^2\Bigr) \cdot |O_A|_2 \\ \nonumber
& \leq & \Bigl( {\rm max}_{\alpha} |A(\alpha)|^2\Bigr) \sqrt{d_A},
\end{eqnarray}
where the last line follows since $\Vert O_A \Vert \leq 1$.

Let $A,B,E$ have Hilbert space dimensions $d_A,d_B,d_E$ respectively (in our particular case, we have $d_B=d_E=D_L)$.  In fact, while the Hilbert space
dimension of $A$ diverges with system size, since the rank of the density matrix on $BE$ is at most $(D'_{L-1})^2$, we can assume $d_A=(D'_{L-1})^2$.
It is not hard to show that ${\rm max}_{\alpha} |A(\alpha)|^2$ is approximately equal to $1/d_A$ times a constant with high probability (i.e., with probability that tends to $1$ as $d_A$ tends to infinity).  To see this, note that the $|A(\alpha)|^2$ are the eigenvalues of the reduced density
matrix on the two sites {\it entering} the isometry $W$.  Each of those two sites is the output of some isometries; call those isometries $V,V'$.
For random choices of $V,V'$, for arbitrary input state to $V \otimes V'$, indeed the output state on the given two sites will have all the eigenvalues
close to $1/d_A$.
So, $|\tilde O_A|_2$ is bounded by a constant times $1/\sqrt{d_A}$, with high probability.
At the same time $|O_B|_2$ is bounded by $\sqrt{d_B}$ if we define the $\ell_2$ norm using the trace on $B$, rather than the trace on $BE$.

We can in fact tighten the bound (\ref{weakbound}) on $|\tilde O_A|_2$, if desired.  Let $\rho$ be the diagonal matrix with entries $|A(\alpha)|^2$.  We have $|\tilde O_A|_2^2={\rm tr}(O_A \rho O_A \rho)$.  Note that ${\rm tr}(O_A \rho O_A \rho)\leq {\rm tr}(O_A \rho \rho^\dagger O_A^\dagger)={\rm tr}(O_A^\dagger O_A \rho^2)$.  For $\Vert O_A \Vert \leq 1$, we have ${\rm tr}(O_A^\dagger O_A \rho^2) \leq {\rm tr}(\rho^2)$.  So, $|\tilde O_A|_2 \leq \sqrt{{\rm tr}(\rho^2)}$.  Note that this is equal to the exponential of minus one-half the $S_2$ entropy of $\rho$.

Define a super-operator ${\cal E}(\ldots)$ by
\be
{\cal E}(O)=\sqrt{\frac{d_B}{d_A}}  {\rm tr}_E(W O W^\dagger).
\ee
This super-operator is a quantum channel multiplied by the scalar $\sqrt{\frac{d_B}{d_A}}$.
Then,
\be
Cor(X:Y) \leq {\rm const.} \times {\rm max}_{\tilde O_A, |\tilde O_A|_2\leq 1} {\rm max}_{O_B, |O_B|_2 \leq 1} \Bigl( {\rm tr}(O_B {\cal E}(\tilde O_A))-
{\rm tr}(O_B {\cal E}(\sqrt{d_A} \rho)) {\rm tr}(\tilde O_A)\Bigr),
\ee
where we have rescaled $\tilde O_A,O_B$ to have $\ell_2$ norm equal to $1$, absorbing factors of $1/\sqrt{d_A}$ and $\sqrt{d_B}$ into ${\cal E}(\ldots)$, and where the constant is present because the bound (before re-scaling) is that $|\tilde O_A|_2$ is bounded by a {\it constant} times $1/\sqrt{d_A}$, with high probability.

We now consider the super-operator ${\cal E}(\ldots)$.  We now consider the case of general $d_A,d_B,d_E$.  The state ${\cal E}(\rho)$, which is the output state of this super-operator for the density matrix as input, may not itself be exactly maximally mixed.  However, it is very close to maximally mixed with high probability if $d_B<<d_A d_E$ and if $\rho$ is close to maximally mixed.  Further, for any traceless operator $O$, we have ${\rm tr}({\cal E}(O))=0$.  Hence, the maximally mixed state is very close to a right-singular vector of ${\cal E}$ if $d_B<<d_A d_E$ and
there is a singular value of ${\cal E}$ very close to $1$.  So, the term 
$-{\rm tr}(O_B {\cal E}(\sqrt{d_A} \rho)) {\rm tr}(\tilde O_A)\Bigr)$ is close to projecting out the largest singular vector of ${\cal E}(\ldots)$.

So, the important quantity for correlations is the magnitude of the second largest singular vector.
Indeed, what we would like to have is that ${\cal E}(\ldots)$ is a non-Hermitian expander (non-Hermitian in that ${\cal E}(\ldots)$ is not Hermitian
viewed as a linear super-operator), meaning that it has one singular value close to $1$ and all others separated from $1$ by a gap.
Calculating the singular values of ${\cal E}(\ldots)$ is likely similar to the calculation in Ref.~\onlinecite{rugqe}, with some additional complications
because we are interested in a very different choice of dimensions.  For one thing, $d_E$ and $d_B$ are comparable here rather than having $d_E<<d_B$.  For another thing,
$d_A \neq d_B$, so the super-operator ${\cal E}(\ldots)$ has a multiplicative prefactor $\sqrt{d_B/d_A}$ compared to a quantum channel.

We leave a proof that it is an expander for a future paper.  However, we give some numerical and analytical evidence.
Let $x=d_B/(d_A d_E)$ and $y=d_A/(d_B d_E)$.
We conjecture that ${\cal E}(\ldots)$ is an expander if $x,y<<1$.
More precisely, what we conjecture is that for a random choice of $W$ with high probability the difference between the largest singular value and $1$ is bounded by some polynomial in $x,y$ and also that the second largest singular value
is bounded by some polynomial in $x,y$.
Note that certainly we do not expect to get an expander if $y \approx 1$.  If $y=1$, then all singular values are equal to $1$.

We can estimate the average over $W$ of the sum of squares of the singular values of ${\cal E}(\ldots)$ using the same techniques
as we used to estimated $E[\exp(-S_2(\ldots))]_W$  previously, as this sum of squares is also a second order polynomial in $W$
and in $\overline W$.  For $d_B=d_E$ and $d_A<<d_B d_E$, one finds that this sum of squares is equal to $d_A$ up to subleading corrections.
The number of non-zero singular values is equal to $d_B^2$ in this case, so that if all singular values (with the exception of the largest) have roughly
the same magnitude, then this magnitude is roughly
\be
\sqrt{d_A/d_B^2}=\sqrt{y}.
\ee

\begin{figure}
\includegraphics[width=12cm]{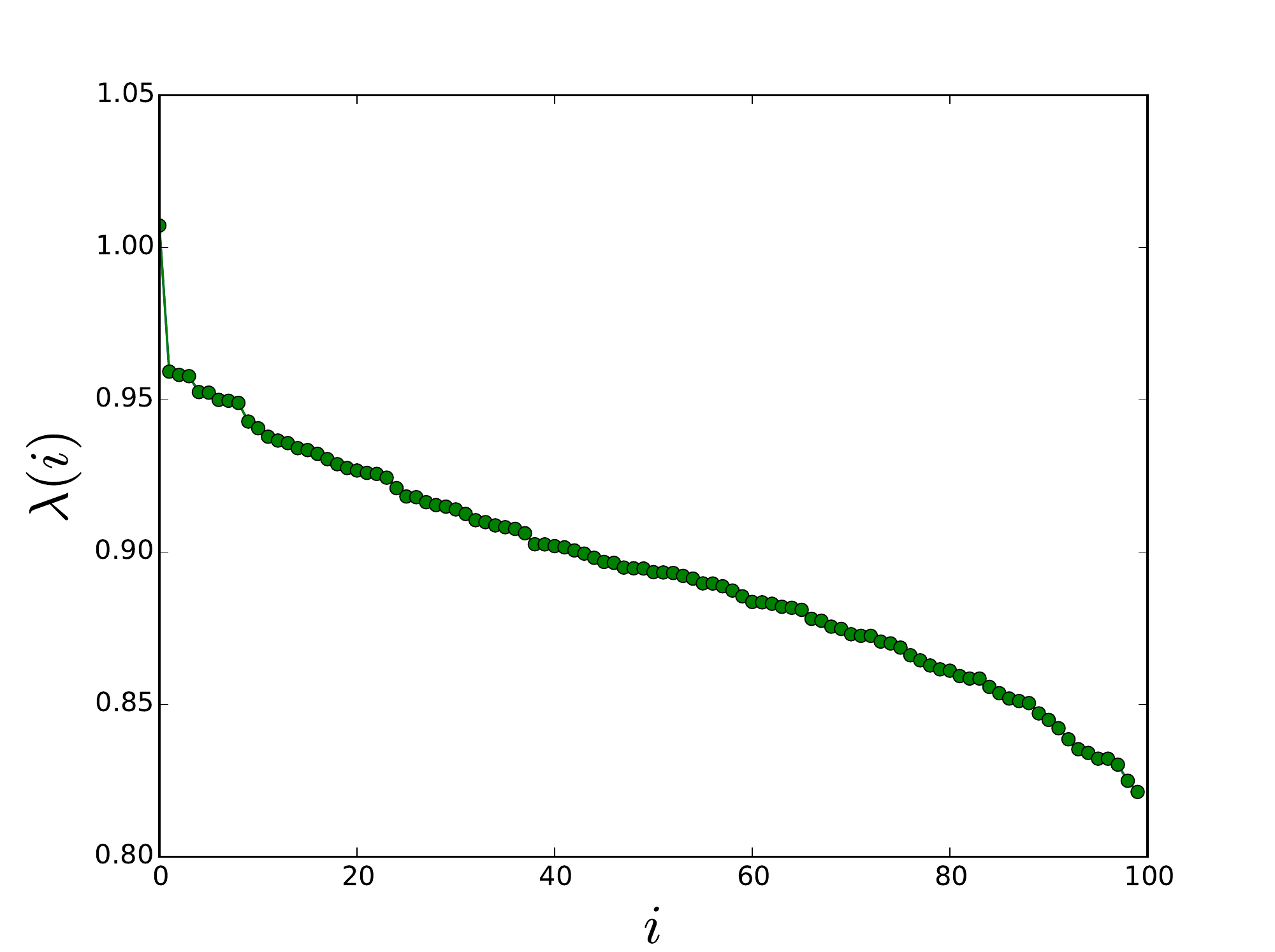}
\caption{Singular values of ${\cal E}(\ldots)$ for a random choice of $W$ for $d_A=80$, $d_B=d_E=10$.  Singular values $\lambda(i)$ are plotted in descending order.  Largest singular value
is equal to $1.0067\ldots$, while next largest singular values are $0.9596\ldots,0.9592\ldots,0.9587\ldots,\ldots$.}
\label{gapfig}
\end{figure}

We have numerically investigated the properties of this super-operator.  First, we observe that qualitatively that there indeed is a gap once $x,y<<1$.
In Fig.~\ref{gapfig}, we show an example with $d_A=80,d_B=d_E=10$.  Even in this case, where $y=0.8$ which is not that small, we observe a distinct gap between the first
singular value and the rest.  We plot the singular values $\lambda(i)$ in descending order as a function of $i$  from $i=0,\ldots,d_B^2-1$.

\begin{figure}
\includegraphics[width=12cm]{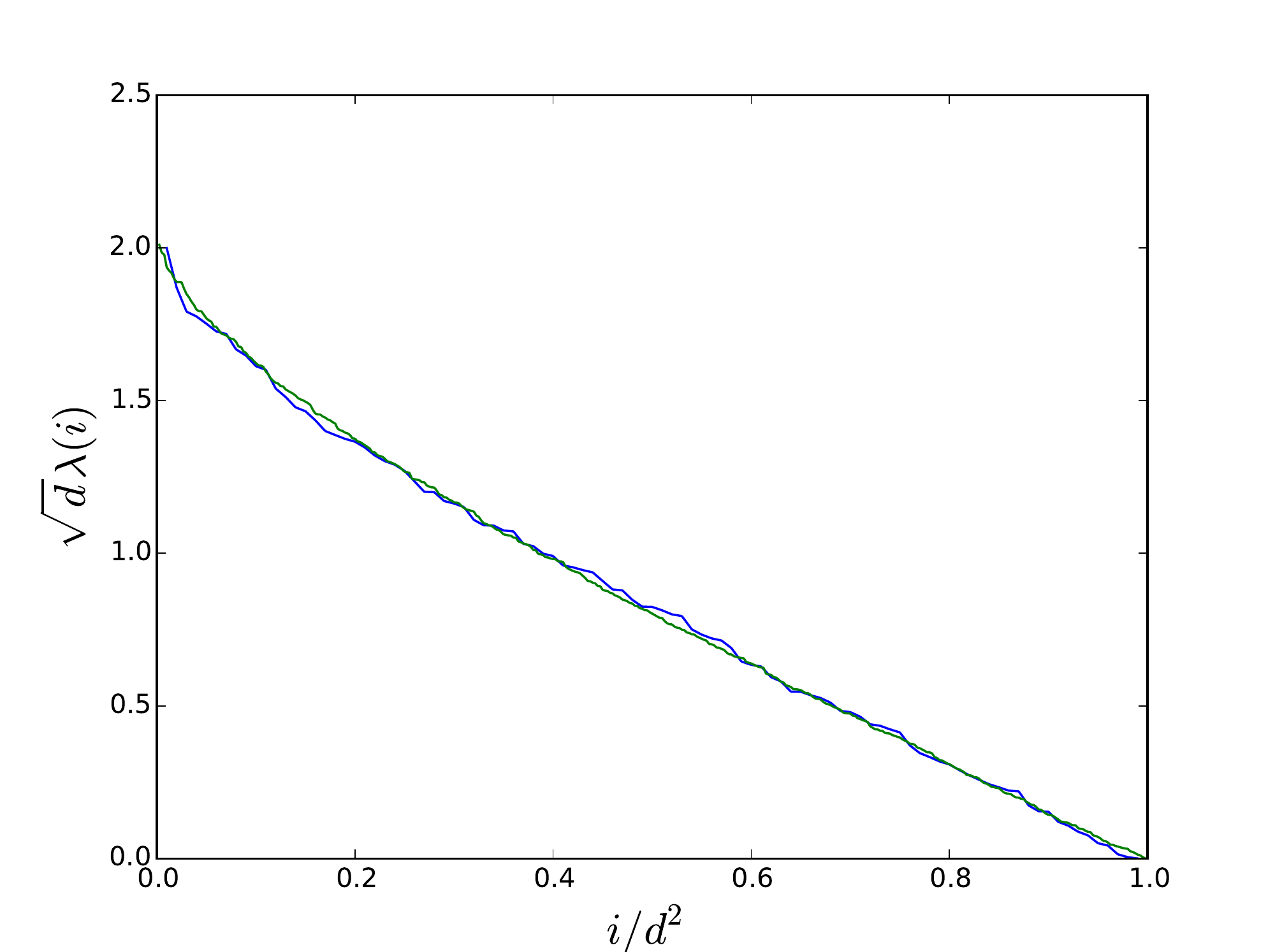}
\caption{Singular values of ${\cal E}(\ldots)$ for two random choices of $W$, one with $d_A=d_B=d_E=d=10$ (shown in blue) and the other with
$d_A=d_B=d_E=d=20$ (shown in green).  $y$-axis shows $\sqrt{d} \lambda(i)$, while $x$-axis shows $i/d^2$.  The largest singular value for each super-operator is not plotted.  The two largest singular values
for the first super-operator are equal to $1.05\ldots,0.645\ldots$, while the two largest second values for the second super-operator are equal to $1.025\ldots,0.456\ldots$.}
\label{collapseFig}
\end{figure}

Next, to test the scaling of the singular values, we first consider the particular case that $d_A=d_B=d_E$.  This is {\it not} the relevant case of interest for the MERA state constructed, however it is still interesting as a way to test the scaling.  In this case, we have $\sqrt{d_A/d_B^2}=1/\sqrt{d_B}$.
What we find is that indeed scaling holds.  We are able to construct a scaling collapse, plotting the singular values $\lambda(i)$ from $i=1,\ldots,d_B^2-1$ in descending order, i.e., not including the leading singular value.  In this plot we plot $\lambda(i)*\sqrt{d_B}$ as a function of $i/d_B^2$.  As shown in Fig.~\ref{collapseFig}, we are
able in this way to almost perfectly collapse curves for different choices of $d_B$.
Further, the collapse holds even for the leading singular values; that is, we have observed that the second largest singular value scales as $1/\sqrt{d_B}$.
So, in this case, we have strong numerical evidence for the polynomial decay as a function of $x,y$.

Before considering the case of interest to us, let us explain why we are interested in having  a polynomial decay of the second largest singular value as a function of $x,y$.
This is due to our desire for
exponential decay of correlations at all length scales, not just for a single $X$ with $Y$ separated from $X$ by one site.  The MERA states used to describe a conformal field theory at criticality display a power law decay of
correlation functions as a function of distance\cite{cftmera}.  To understand this polynomial decay, consider a correlation of two operators $O_i,O_j$ supported on single sites, $i,j$.  Then one can iteratively map an operator such as $O_i$ or $O_j$ into an operator at higher levels of the MERA.  This map is a linear map; it is in fact related to the adjoint of a super-operator such as ${\cal E}(\ldots)$ that we consider; the map in Ref.~\onlinecite{cftmera} is regarded as moving operators up to higher levels of the MERA rather than, as we have described it, moving states to lower levels of the MERA.  To move up one level in the MERA, one must apply two super-operators, as each level in the MERA corresponds to two isometries $V,W$.  This linear map leads to an exponential decay of the difference between the operator and the identity operator as a function of level in the usual
MERA states; since the number of levels between $i,j$ is logarithmic in $i,j$, this leads to a polynomial decay.  The reason for the exponential decay is that
in such MERAs, the isometries are taken in a scale-invariant fashion, so that they are the same at all levels (or all except the bottom few levels) and so the super-operator has a fixed gap
to the second largest singular value at all levels.
In our MERA, however, the isometries
change with level.  Thus, we hope that the decay when moving from one level to the next will be polynomial in $x,y$.  Since $y \approx \exp(-\epsilon 2^{L-k})$ for isometries in $W_k$,
a polynomial decay in the smallest $y$ (which occurs at the highest level, giving a $y$ which is exponentially small in the spacing between sites) will lead to an exponential decay in $i-j$.

One complication in this is that when we map an operator on a single site $i$ of the MERA to higher levels of the MERA, result is no longer an operator supported on a single site.
However, the so-called {\it causal cone} of such an operator (i.e., the support of the operator after it is mapped to higher levels of the MERA; this support is the same as the set of sites which have $i$
in their light-cone as we have defined the light-cone) does not consist of a single site at each level.  Rather, the causal cone consists of some small number of sites\cite{cftmera}, depending upon the exact MERA chosen.
However, it seems likely that, since we are considering an $\ell_2$ norm, if we can show a gap in the singular values of the super-operator corresponding to the map of a single site operator
upwards by one level of the MERA, it will also be possible to show a gap in the map of an operator supported on some small number of sites,
as the $\ell_2$ norm does have the nice property that the singular values of a product of super-operators can be determined from the singular values of the individual super-operators.  If
we instead worked with $\ell_\infty$ norms, there would be difficult multiplicativity questions that would arise and perhaps having a bound in the $\ell_\infty \rightarrow \ell_\infty$ norm of a pair of super-operators
would not help in bounding the $\ell_\infty \rightarrow \ell_\infty$ norm of the product.
In this way, we conjecture that it will be possible to show at least an exponential decay of $Cor(X:Y)$ for distances sufficiently large compared to the diameter of $X$ and the diameter of $Y$.

A more difficult question is whether we can show an exponential decay even if the diameters of $X,Y$ are large compared to the distance between $X,Y$.  We conjecture that this will also hold.  We take an operator $O_X$ and apply the super-operator ${\cal E}(\ldots)^\dagger$ to map $O_X$ upward in the MERA and similarly map $O_Y$ and apply this process repeatedly until $X,Y$ meet.
The intuitive idea is that at every step of this process we consider the site $i$ at the leftmost edge of $X$ and we decompose the operator $O_X$ on $X$ into a sum of two terms, $O_X^0+O_X^\perp$, where $O_X^0$ is the identity operator on $i$ tensored with some other operator on the rest of $X$, while $O_X^\perp$ vanishes after tracing over $i$.
The site $i$ is one of two sites output from some given isometry.  Assume that the other site output from that isometry is to the left of $i$ so that it is not in $X$; in this case we say that ``a site is traced over at the left end".  Note that it is not necessary that a site be traced over on a given step; for example, if $X$ consists of two sites which are output from the same isometry,
then no site is traced over.  However, if a site is not traced over on a given step, then a site must be traced over at the next step.
So, suppose that a site is traced over.
Let ${\cal E}(\ldots)$ be the super-operator associated with this isometry and tracing over the site $i-1$.
We would then use the bound on singular values of the super-operator ${\cal E}(\ldots)$ to show that the $\ell_2$ norm of $O_X^\perp$ decays by an amount $\exp(-{\rm const.} \times \epsilon 2^{L-k})$ after applying the super-operator ${\cal E}(\ldots)^\dagger$ to map it to an operator higher in the MERA, while $O_X^0$ maps to an operator with increased separation between $X$ and $Y$.  
In this manner, we conjecture that at some level $k$ , with $k \sim \log(l)$, we must have a decay in $\ell_2$ norm by $\exp(-{\rm const.} \times \epsilon 2^{L-k})$.

\begin{figure}
\includegraphics[width=12cm]{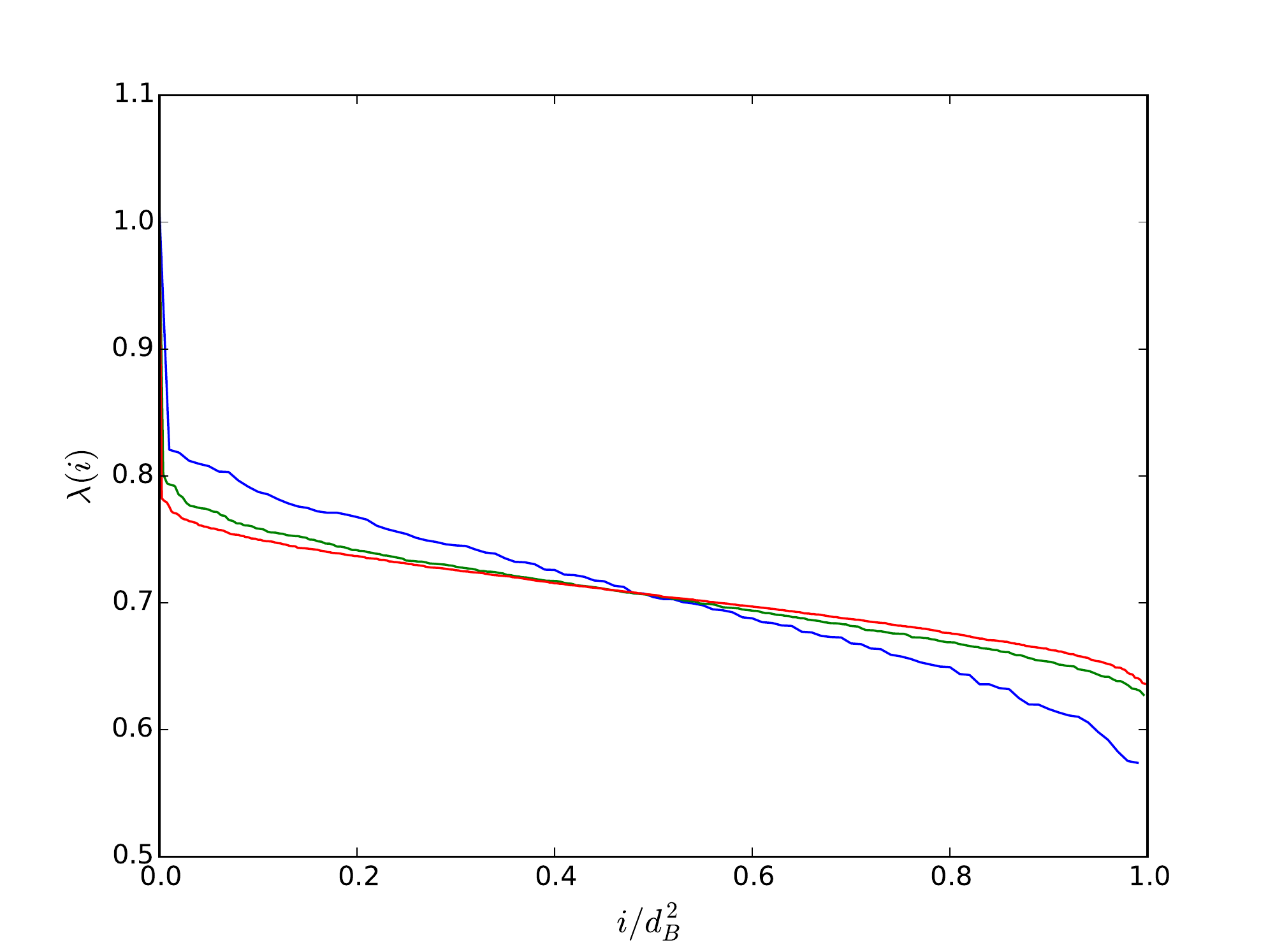}
\caption{Singular values of ${\cal E}(\ldots)$ for three random choices of $W$, one with $d_A=50$, $d_B=d_E=d=10$ (shown in blue), $d_A=128$, $d_B=d_E=16$ (shown in green), and
$d_A=200, d_B=d_E=20$ (shown in red).  $y$-axis shows $\lambda(i)$, while $x$-axis shows $i/d^2$.}
\label{nocollapseFig}
\end{figure}

\begin{figure}
\includegraphics[width=12cm]{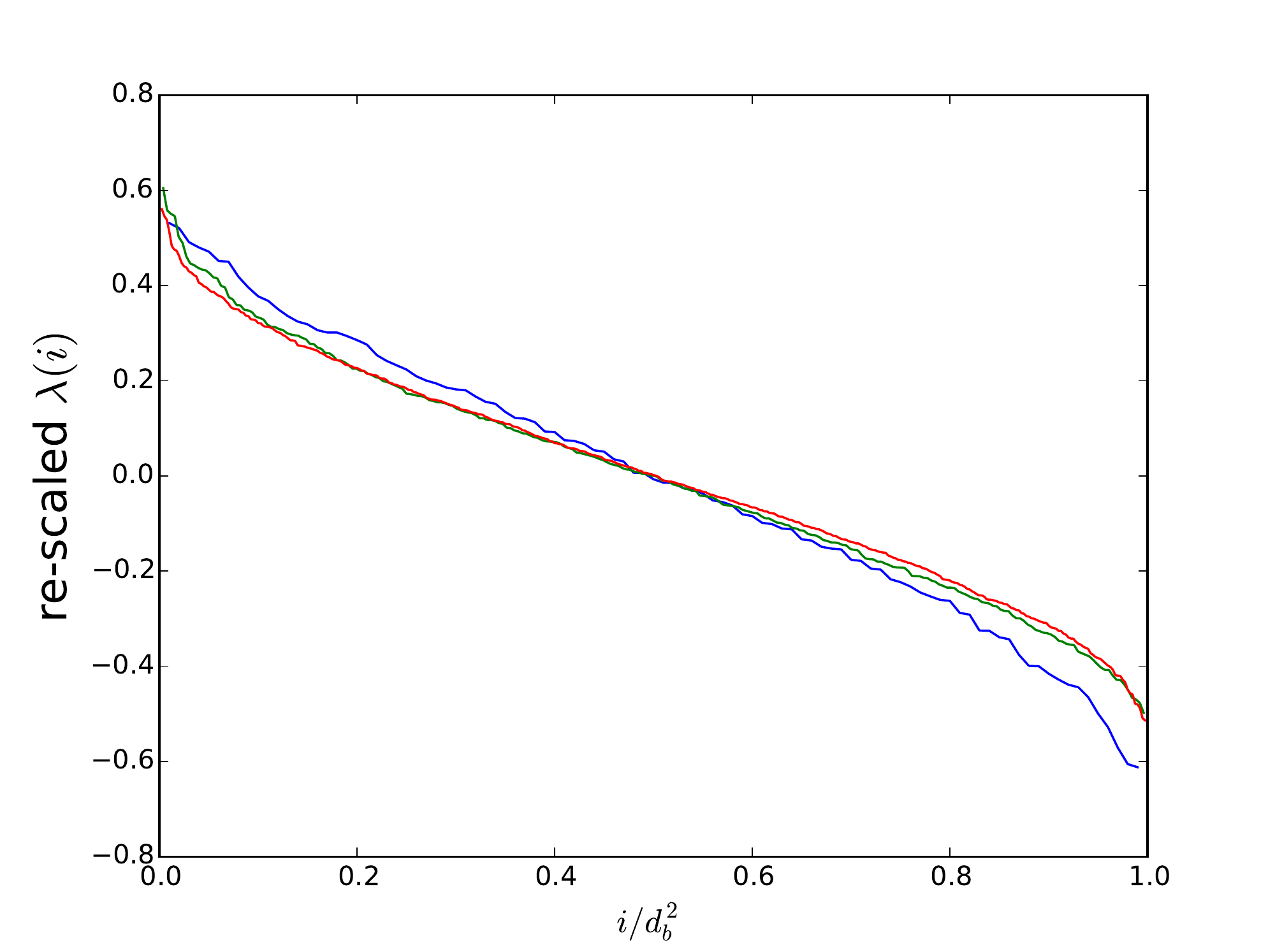}
\caption{Re-scaled singular values for same three channels as in Fig.~\ref{nocollapseFig} for $i=1,\ldots,d_B^2-1$.}
\label{coll3Fig}
\end{figure}

When we turn to the case of interest to us, with $d_A>>d_B$ but $y<<1$, we do not find a clear scaling collapse.  In this case, since $x<<y$, we might hope
that the scaling collapse would hold with two different super-operators with the same $y$.  In Fig.~\ref{nocollapseFig}, we see that this is not the case for three different super-operators with $y=1/2$
for both and $d_B=d_E=10,16,20$.   It is possible, however, that for large enough $d_B$ at fixed $y$ the singular values will eventually collapse on each other; the curves are becoming flatter with
increasing $d_B$ suggesting that this may happen.  If such a collapse happens for large $d_B$ for the entire curve, then indeed the second largest singular value must be proportional to $\sqrt{y}$ for large $d_B$.  Even if there are corrections to this which vanish polynomially in $d_B$, this would still suffice.  Some evidence for a collapse is shown in Fig.~\ref{coll3Fig}.
Here we show an attempt to collapse the three curves by rescaling $(\lambda(i)-{\rm const.}) d_B^{\alpha}$, where the constant $0.706\ldots$ is chosen to match the approximate
crossing point of the curves and $\alpha=2/3$ was chosen after some experimentation.  Good collapse is seen between the curves with $d_B=16,20$, while the curve with $d_B=10$ does not collapse as well, especially for large $i$.

\section{Discussion}
While this work was in progress, another work constructed a MERA state for which the entanglement entropy was exactly given by the minimum length of curves cutting through the MERA network\cite{adsmera}.  There are two main differences in the type of states constructed.  First, we used random tensors,
instead of the perfect tensors used there.  Second, we considered a very different set of choices of dimensions at different levels of the MERA, in our goal of constructing a state with high entanglement and low correlations.  These two choices may have some interpretation in the language of holography and quantum gravity as follows.  The different choice of dimensions may correspond to some different choice of geometry in the bulk space, rather than
an $AdS$ geometry.

The choice of random tensors, however, might be interpretable in terms of quantum fluctuations in the bulk geometry: instead of the entanglement entropy being exactly expressed in terms of a single curve cutting through the MERA, the optimum reduction sequence (note that each reduction sequence corresponds uniquely to a curve) gives only upper and lower bounds on the expected entanglement entropy, with a possible logarithmic difference between those results.  However, the expected exponential of minus the $S_2$ Renyi entropy,
$E[\exp(-S_2(\ldots))]$, can be exactly expressed as a {\it sum} over reduction sequences (or curves).  This difference between a minimization
and a sum is reminiscent of the difference between classical and quantum mechanics (least action path compared to path integral).  If the dimensions $D_k,D'_k$ become large (and importantly also the differences between certain sums of the $D_k,D'_k$ become large), then the sum becomes dominated by a single curve.  This is perhaps reminiscent of the fact that certain random matrix theories can be interpreted as a sum over random surfaces, with the
limit of large matrix size in the random matrix theory involving a sum only over a single genus; our theory is a more general kind of random matrix theory, but  
perhaps something similar happens.

Finally, the reader might note that we only prove results about the expectation value of the entanglement entropy, rather than proving results about
the entanglement entropy for a specific choice of isometries in the MERA.  
For example, lemma \ref{intentlemma} only lower bounds the expectation value of the entanglement entropy for
intervals of length $l$.  The reader might wonder: is there a specific choice of isometries for
which for all intervals of length $l$, the entanglement entropy is within some constant factor of its expectation value?  In this paper,
we did not worry at all about trying to prove such results.  However, we briefly mention some possible ways to try to do this.
One might, for example, try to use concentration of measure arguments to estimate fluctuations about the
average.  This could perhaps show that the probability of a ``bad event", such as low entanglement entropy (or perhaps long correlation length, if indeed it is true that the state is short-range correlated as we conjecture), is exponentially small in dimension.
This approach has the downside that the system size is exponentially large in $D_L$, so that even if a bad event is exponentially unlikely in any particular
part of the system, it may be likely to occur somewhere.
To resolve this issue, one might try to use the Lovasz local lemma in some way: it might be possible to show that the event that the entanglement entropy of
some given interval $[i,j]$ was small was independent of the event that the entanglement entropy of some other interval $[i',j']$ was small if $|i-i'|,|j-j'|$ are sufficiently large.  Or, more simply stated: perhaps if a bad event occurs locally, one might resample those isometries and leave the other isometries unchanged.
Perhaps another approach to avoiding having bad events occur somewhere is to reduce the amount of randomness: rather than choosing all isometries independently at random, one might instead take
all isometries $W$ at a given level to be the same and sample that isometry at random, independently for each level, and similarly take all $V$ at a given level to be the same.  This approach has the downside that
it complicates the calculations of the entanglement entropy.  For example, if we consider 
$S_2(\sigma(k)_{[i,j]}$ and $i \md 2=0$ and $j \md 2=1$, then $\exp(-S_2(\ldots))$ is now a fourth order polynomial in $W$ and $\overline W$, where
$W$ is the isometry at the given level of the MERA.  This leads to additional terms in the equation for $S_2$, beyond those in Eq.~(\ref{exprecursigma}).
These extra terms likely do not change the result that we have found for the mutual information, however.

One simple way to reduce the randomness without complicating the calculation of the asymptotic behavior of the entanglement entropy
is to choose the isometries at each level of the MERA to repeat with some sequence.  That is, if we consider the isometry $W$ in some given level of the MERA, if this isometry is a product of $n$ isometries on pairs of sites, rather than choosing them all independently as done in this paper, and rather than choosing them all the same, one could choose them so
that $W_1,...,W_a$ are sampled independently for some $a$, and then have the sequence repeat so that $W_i=W_{i-a}$.
In this way, if we calculate entanglement entropy of an interval short compared to $a$, we find the same Eqs.~(\ref{exprecursigma},\ref{exprecurtau}).
We keep $a$ the same at every level; then, a large interval of some length $l$ would have additional terms present at the lower levels, but once one reached a level
of the MERA of order $\log_2(l)$, then we find the same  Eqs.~(\ref{exprecursigma},\ref{exprecurtau}); note that it is at such a level that
the dominant contributions to the entanglement entropy occur and so one finds the same results as in lemmas \ref{intentlemma},\ref{milemma} for the asymptotic behavior.
We leave these questions aside, however, until a proof of the correlation decay is given.

As a final remark, one may modify the state by changing the recursion relations (\ref{apprecur1},\ref{apprecur2}) by replacing the
factor $2^{L-k}$ by $(2^{L-k})^{\kappa}$ for an exponent $\kappa$.  Having done this, for any $\kappa<1$, for sufficiently small
$\epsilon$, one can take $L$ arbitrarily large (i.e., $L$ is no longer restricted to be of order $1/\epsilon$) and have $\log(D_{k}),\log(D'_k)$ roughly
proportional to $2^{L-k}$ (the factor $(2^{L-k})^{\kappa}$ becomes negligibly small).  In this manner, it seems likely that the resulting
state will combine a {\it volume} law for entanglement entropy with almost exponentially decaying correlation functions (correlation between two regions separated by $l$ sites proportional to $\exp(-l^\kappa/{\rm const.})$ for some constant).  Generalizing this to higher dimensional
MERA states\cite{hdmera}, we conjecture that one can obtain MERA states in $d$ spatial dimensions with volume law entanglement
and correlations decaying as $\exp(-l^\kappa/{\rm const.})$ for any $\kappa<d$ (in particular, for $d=2$ it seems that one can obtain super-exponential correlation decay and volume law entanglement).

{\it Acknowledgments---} I thank Fernando Brandao for explaining Ref.~\onlinecite{bh} and for many useful discussions especially on the generalizations discussed in the last paragraph.  I thank C. King for useful comments on multiplicativity of norms for
super-operators.

\end{document}